\documentclass[10pt,final]{IEEEtran}

\usepackage{amsthm}
\usepackage{amsmath}
\usepackage{latexsym}
\usepackage{graphicx}
\usepackage{bbding}
\usepackage{indentfirst}
\usepackage{algorithm}
\usepackage{setspace}
\usepackage{float}
\usepackage{epstopdf}
\usepackage{amssymb}
\usepackage{amsfonts}
\usepackage{enumerate}
\usepackage{multicol}
\usepackage{color}
\usepackage{multirow}
\usepackage{bm}
\usepackage{amssymb}
\usepackage{stfloats}
\usepackage{epstopdf}
\usepackage{threeparttable}
\usepackage{epstopdf}
\usepackage{threeparttable}
\usepackage{subfigure}
\usepackage{makecell}
\usepackage{cite}
\usepackage{cancel} 
\usepackage{multirow}
\usepackage{array}     
\usepackage{hyperref}  
\usepackage{graphicx}  
\usepackage{float}  
\usepackage{subfigure}  
\usepackage{amsmath,amssymb}       
\usepackage{booktabs}              
\usepackage{graphicx}              
\usepackage{caption}               
\usepackage{float}                 
\usepackage{array,multirow}
\newcolumntype{C}[1]{>{\centering\arraybackslash}m{#1}}
\newcolumntype{L}[1]{>{\raggedright\arraybackslash\setlength\parindent{0pt}}m{#1}}

\usepackage{subcaption} 
\IEEEoverridecommandlockouts
\allowdisplaybreaks[4]

\usepackage{cite}
\usepackage{amsmath,amssymb,amsfonts}
\usepackage{algorithm}  
\usepackage{algpseudocode}  
\usepackage{algpseudocode}
\usepackage{graphicx}
\usepackage{epstopdf}
\usepackage{textcomp}
\usepackage{xcolor}
\usepackage{bm}
\usepackage{setspace}
\usepackage{tabularx}
\usepackage{booktabs}
\usepackage{array}
\usepackage{longtable}

\begin{document}
\title{Intelligent Reflecting Surfaces for Integrated Sensing and Communications: From System Coexistence to Networked Mutualism}

\author{{Qingqing Wu,~\IEEEmembership{Senior Member,~IEEE}, Qiaoyan Peng, Ziheng Zhang, Xiaodan Shao, Yang Liu, Yifan Jiang, \\ Yapeng Zhao, Yanze Zhu, Yilong Chen, Zixiang Ren, Jie Xu,~\IEEEmembership{Fellow,~IEEE}, Wen Chen,~\IEEEmembership{Senior Member,~IEEE}, \\Rui Zhang,~\IEEEmembership{Fellow,~IEEE}}
	\thanks{Q. Wu, Z. Zhang, Y. Zhu, and W. Chen are with the Department of Electronic Engineering, Shanghai Jiao Tong University, Shanghai 200240, China (email: qingqingwu@sjtu.edu.cn; zhangziheng@sjtu.edu.cn; yanzezhu@sjtu.edu.cn; wenchen@sjtu.edu.cn).
    Q. Peng and Y. Jiang are with the Department of Electronic Engineering, Shanghai Jiao Tong University, Shanghai 200240, China, and also with the State Key Laboratory of Internet of Things for Smart City, University of Macau, Macao SAR, China (email: qiaoyan.peng@connect.um.edu.mo; yc27495@um.edu.mo).
    X. Shao is with Department of Electrical and Computer Engineering, University of Waterloo, Waterloo, ON N2L 3G1, Canada (e-mail: x6shao@uwaterloo.ca).
    Y. Liu is with the School of Information and Communication Engineering, Dalian University of Technology, Dalian 116024, China (e-mail: yangliu\_613@dlut.edu.cn).
    Y. Zhao is with College of Electronics and Information Engineering, Shenzhen University, Shenzhen 518060, China (email: ypzhao@szu.edu.cn).
    Y. Chen, Z. Ren, and J. Xu are with School of Science and Engineering, Shenzhen Future Network of Intelligence Institute (FNii-Shenzhen), and Guangdong Provincial Key Laboratory of Future Networks of Intelligence, The Chinese University of Hong Kong, Shenzhen, Guangdong 518172, China (e-mail: yilongchen@link.cuhk.edu.cn; rzx66@mail.ustc.edu.cn; xujie@cuhk.edu.cn).
    R. Zhang is with the Department of Electrical and Computer Engineering, National University of Singapore, Singapore 117583 (e-mail: elezhang@nus.edu.sg).
    }
}

\maketitle
\IEEEpeerreviewmaketitle

\begin{abstract}
The rapid development of sixth‑generation (6G) wireless networks requires seamless integration of communication and sensing to support ubiquitous intelligence and real‑time, high‑reliability applications. Integrated sensing and communication (ISAC) has emerged as a key solution for achieving this convergence, offering joint utilization of spectral, hardware, and computing resources. However, realizing high‑performance ISAC remains challenging due to environmental line‑of‑sight (LoS) blockage, limited spatial resolution, and the inherent coverage asymmetry and resource coupling between sensing and communication. Intelligent reflecting surfaces (IRSs), featuring low‑cost, energy‑efficient, and programmable electromagnetic reconfiguration, provide a promising solution to overcome these limitations. This article presents a comprehensive overview of IRS‑aided wireless sensing and ISAC technologies, including IRS architectures, target detection and estimation techniques, beamforming designs, and performance metrics. It further explores IRS‑enabled new opportunities for more efficient performance balancing, coexistence, and networking in ISAC systems, focuses on current design bottlenecks, and outlines future research directions. This article aims to offer a unified design framework that guides the development of practical and scalable IRS‑aided ISAC systems for the next‑generation wireless network.
\end{abstract}


\section{Introduction}

\subsection{Background}

The forthcoming sixth-generation (6G) wireless communication network is anticipated to facilitate a broader range of services and applications with enhanced communication and sensing capabilities compared to the current commercial fifth-generation (5G) network \cite{c1}. 
For instance, a peak data rate of 200 gigabits per second and a latency of 0.1-1 millisecond are expected to support numerous emerging real-time applications, including extended reality, digital twins, autonomous driving, and remote surgery \cite{c2,c3}. 
Furthermore, building upon the existing large-scale and multidimensional deployment of wireless network infrastructures, 6G holds significant potential for flexible and high rate wireless link establishment and seamless wireless coverage. 
Such extraordinary capabilities are essential for promoting new application domains such as pervasive intelligence and low-altitude economy (LAE) \cite{c4}.

Integrated sensing and communication (ISAC), identified as one of the 6G key usage scenarios and technologies, has garnered substantial research interest from academia and industries worldwide \cite{c1}. 
Specifically, ISAC refers to the integrated usage of hardware, wireless resources, and signal processing techniques to facilitate both sensing and communication functions \cite{c6}. 
By effectively integrating the dual capabilities and leveraging their mutual assistance, ISAC can significantly enhance both sensing and communication performances to unprecedented levels in various scenarios, including Internet of Things (IoT), vehicle-to-everything (V2X), and unmanned aerial vehicle (UAV) networks. 
A promising prospect of ISAC is to achieve highly accurate target detection, localization, tracking, and recognition that have evolved from existing wireless network architectures, especially cellular networks.

In recent years, the upgraded capability of reconfiguring wireless channels has become the most important technical breakthrough in wireless network design. 
As a representative enabler, intelligent reflecting surfaces (IRSs) can flexibly reflect impinging wireless signals towards anticipated directions/positions and significantly enhance the reflected signal strength with a substantial beamforming gain \cite{c2,wu2021-intelligent, chen_iot}. 
In addition, IRS can be conveniently deployed on either side of the transmitter, environment, or receiver, with low hardware cost and high energy efficiency. 
So far, IRS has undergone considerable development from theoretical concepts to experimental verification \cite{direnzo2020-smart,wu2019-intelligent,c2,wu2025IRS}. 
Therefore, it is envisioned that the IRS can provide new and rich degrees of freedom (DoFs) for ISAC system design, thereby greatly enhancing both sensing and communication performance in future 6G networks.

\subsection{Challenges for Wireless Sensing and ISAC}
The anticipated high performance of 6G networks poses considerable technical challenges for wireless sensing and communication design. To be specific, it is challenging to boost the sensing performance of existing wireless network infrastructure, due to the following main issues:

\subsubsection{Non-availability of line-of-sight (LoS) links}
LoS links are typically vital to wireless sensing because not only the propagation path loss of sensing signals can be reduced but also the target properties (e.g., angle, distance, and velocity) can be efficiently obtained via wireless signal reflection in LoS channels.
However, LoS channel conditions are usually not satisfied because of complicated environmental factors such as widely distributed blockages, obstacles, and unexpected scatterers.
In many cases, such factors cause severe signal attenuation and unfavorable multipath interference, further degrading the sensing performance.

\subsubsection{Limited far-field (FF) spatial resolution}
The spatial resolution is one of the most crucial sensing performance metrics for target detection, estimation, and identification.
In general, the spatial resolution in FF conditions mainly depends on the system resources, such as the signal waveform, bandwidth, and beamwidth \cite{9737357}.
Unfortunately, such resources are limited owing to practical hardware, spectrum, and energy constraints.
Therefore, it is challenging to achieve a universally high FF spatial resolution, given limited resources.

\subsubsection{Limited sensing range}
The sensing range refers to the area where targets can be effectively sensed, which significantly influences sensing coverage performance.
However, the sensing range of existing wireless systems is limited by blockage and signal propagation path loss.
For example, in millimeter-wave (mmWave) and terahertz (THz) systems, obstacles and long transmission distances can severely degrade the signal strength, leading to considerable declines in sensing performance. 
In addition, the existence of environmental clutter and signal interference makes it challenging to provide a reliable sensing range \cite{YYNiu2025IOT}.

In addition to the above inherent issues of wireless sensing, there is an urgent need to address the new challenges of ISAC systems.
Two main challenges are outlined as follows:
\setcounter{subsubsection}{0}
\subsubsection{Asymmetric communication/sensing coverage}
The sensing and communication coverage ranges are asymmetric owing to different signal propagation processes and service requirements \cite{Net26_CHuang_TWC2025}.
For instance, in a monostatic ISAC system, communication signals generally experience one-way path loss from the transmitter to the receiver, whereas sensing is based on signals through round-trip reflection, thus resulting in significantly shorter sensing ranges compared to communication ranges.
Such asymmetry poses challenges in characterizing sensing and communication coverage as well as achieving a flexible trade-off between sensing and communication performance.

\subsubsection{Communication/sensing resource sharing}
Owing to the limited system hardware and radio resources, it is challenging to design resource sharing schemes for coexistence and flexible performance balancing between sensing and communication, such as spectrum sharing for mutual interference mitigation and bandwidth allocation satisfying both sensing and communication requirements. 
Additionally, resource sharing for reciprocity between sensing and communication is a new challenge for ISAC system designs. 
More fundamentally, it is non-trivial to characterize and achieve the maximum sensing and communication performance trade-off bound of ISAC systems. 

\subsection{IRS's New Opportunities}
As an emerging technology for improving the wireless environment, IRS has brought significant system performance gains and abundant new design DoFs in wireless communication networks. 
Similarly, the IRS also offers substantial opportunities to address the aforementioned challenges and enhance the performance of wireless sensing, elaborated as follows:
\subsubsection{Favorable channel establishment}
    IRS can be applied in non-line-of-sight (NLoS) scenarios to achieve signal propagation in cascaded LoS target-IRS and IRS-radar channels by redirecting the incident signals, thereby forming a virtual LoS channel. 
    For example, IRS can be utilized to facilitate around-the-corner target detection, where the direct channel between the radar and the target is blocked \cite{LoS-1}. 
    In addition, applying IRS in sensing systems introduces additional controllable paths for signal propagation, which can be exploited to improve sensing performance and mitigate interference.
    Thus, the number of exploitable channels favorable for sensing is considerably increased, and many issues caused by NLoS channel conditions can be resolved. 

\subsubsection{Spatial resolution enhancement}
    By providing substantial additional resources for wireless sensing systems, IRS is promising to improve the spatial resolution to an unprecedentedly high level, especially for FF target detection. 
    For instance, the IRS's large aperture can generate narrow beams to distinguish neighboring targets, thus significantly improving the angular resolution \cite{Resolution-1}. 
    Additionally, the low cost and power consumption of IRS make large-scale IRS deployment practical, therefore, it can achieve pervasive high-resolution target detection and identification. 
    
\subsubsection{Sensing range improvement}
    The large beamforming gain enabled by the large number of IRS reflecting elements can compensate for the severe path loss and thus achieve a larger sensing range. 
    Besides, the IRS's capability to redirect signals can facilitate wireless signal coverage for dead zones, thereby expanding the sensing range. 
    Moreover, the cascaded links among multiple IRSs enable a larger distance for detecting targets than conventional systems without deploying IRS.
    
\subsubsection{Target radar cross section (RCS) improvement}
    Thanks to the IRS deployment flexibility, an IRS with a large aperture can be mounted on targets with a small RCS, such as UAVs and vehicles. 
    By this means, the target RCS is increased, and the target reflectivity is improved and controllable, which offers additional design DoFs for system performance enhancement.

In addition to facilitating wireless sensing, the use of IRSs can further benefit ISAC systems from the following perspectives:
\setcounter{subsubsection}{0}
\subsubsection{Communication/sensing balancing}
    Thanks to the flexible tunability of the reflecting elements, IRS can achieve a flexible balance between communication and sensing performance. 
    For example, the communication and sensing signal‑to‑noise ratio (SNR) can be flexibly balanced by optimizing the IRS's reflection coefficients or by dividing the IRS into subarrays for different functions. 
    This capability to balance communication and sensing performance is promising for addressing the intrinsic asymmetric communication/sensing coverage issue. 
\subsubsection{Flexible resource allocation}
    Depending on the specific application, IRS's resources can be flexibly allocated to boost communication and sensing performance simultaneously or to achieve their reciprocity, including the reflection coefficients, number of active/passive elements, shape, operating frequency bands, and deployment location and orientation. 
    Additionally, IRS resources can be allocated separately to mitigate the interference between communication and sensing by generating dedicated communication and sensing beams.

\subsection{Related Works and Contributions}

Some existing overviews \cite{o4, o2, 2023_Asif2, o8, o5, o6, 10422881, o1, o3}, tutorials \cite{t1}, and surveys \cite{s1, s5} have investigated the opportunities and design challenges of IRS-aided sensing/ISAC. 
Specifically, many studies focused on IRS-aided sensing and the coexistence of communication and sensing \cite{o4,o2,2023_Asif2,o8,o6,10422881,t1,o5,s1,s5}. 
Additionally, some technical principles of mutualism in IRS-aided ISAC systems were investigated in \cite{o1}, and some existing studies on multi-IRS-aided networked ISAC were surveyed in \cite{o3}.
Compared to these related studies, this article provides a comprehensive survey of research on not only wireless system architectures and signal processing techniques for IRS-aided sensing/ISAC, but also mutualism and networked designs for IRS-aided ISAC. 
Particularly, this survey first overviews the IRS architectures, techniques for target detection/estimation/tracking, beamforming designs, and performance metrics for sensing. 
Then, the opportunities, solutions, and challenges of IRS-aided ISAC are surveyed from the perspectives of coexistence, reciprocity, and the networked ISAC paradigm, respectively.
The aim of this article is to provide a useful guide for inspiring more innovative, effective, and practical designs facilitating IRS-aided sensing/ISAC in future wireless networks. 

The rest of this article is organized as follows. 
Section II comprehensively overviews the architectures, target detection/parameter estimation/tracking algorithms, beamforming designs, and performance metrics of IRS-aided sensing. 
Sections III, IV, and V discuss the applications, design issues, and optimization algorithms of IRSs for coexistence, mutualism, and the networked paradigm of ISAC, respectively. 
In Section VI, promising future directions of IRS-aided sensing/ISAC research are discussed. 
Finally, Section VII concludes this article. 




\section{IRS-aided sensing}
The sensing coverage is often constrained by severe path loss in the base station (BS)-IRS-target-IRS-BS cascaded echo link, which limits the effectiveness of target detection and parameter estimation. By integrating IRS, wireless system can effectively improve the sensing performance and reduce both hardware cost and energy consumption compared with deploying dense BSs \cite{10540249,10740590}. This section provides an extensive review of IRS sensing in terms of IRS sensing architecture, sensing performance metrics, detection and estimation algorithms, as well as beamforming design.


\subsection{IRS Sensing Architecture}
\begin{figure*}[!t]
	\centering
	\includegraphics[width=0.8\textwidth]{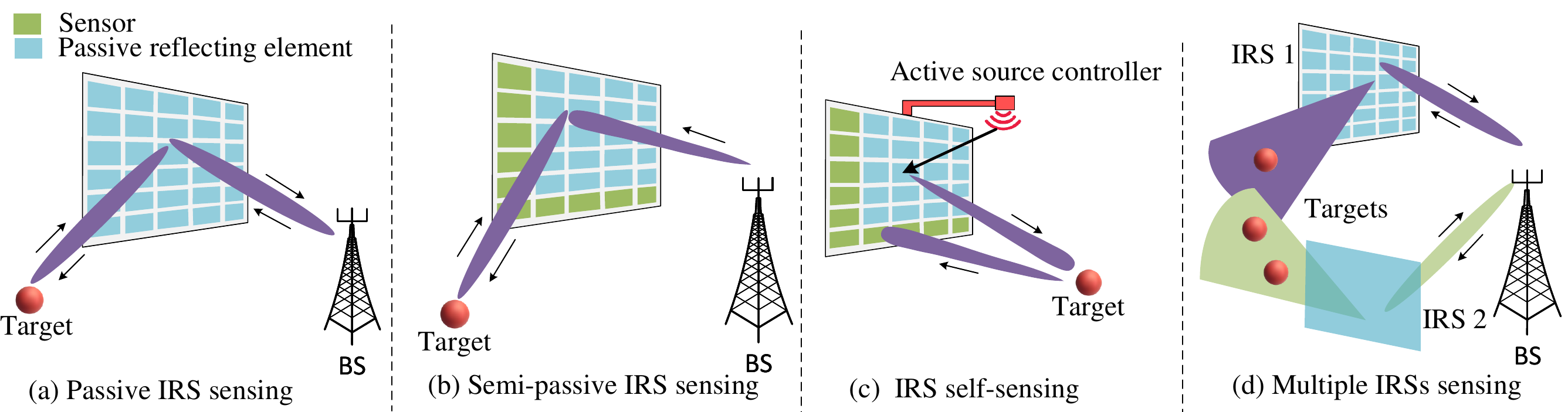}
	\caption{Environmentally-deployed IRS for sensing.}
	\label{environment}
    \vspace{-5pt}
\end{figure*}

Prospective IRS architectures are essential for overcoming the coverage bottleneck in wireless sensing systems. As shown in Fig. \ref{environment}-Fig. \ref{advanced}, three typical IRS-enabled wireless sensing architectures can be identified: the environmentally-deployed IRS, the target-mounted IRS, and advanced IRS architectures.

\subsubsection{Environmentally-deployed IRS}
As shown in Fig. \ref{environment}, an IRS can be strategically placed within the surrounding environment to enhance sensing coverage and signal quality. In this category, there are four main approaches: IRS passive sensing, IRS semi-passive sensing, IRS self-sensing, and multi-IRS-aided sensing, each leveraging different levels of IRS functionality and hardware complexity to meet diverse sensing requirements.

First, for the passive IRS sensing without dedicated sensors, the target sensing is performed at the BS based on the echo signals through the BS-IRS-target-IRS-BS link \cite{shao2022target,10422881} (see Fig. \ref{environment}(a)). This method enables the acquisition of more informative measurements and enhances the accuracy of various sensing applications.
Passive IRS sensing divides into two classes: low‑frequency sensing and high‑frequency sensing. In the low‑frequency regime, the authors in \cite{10138058} investigated IRS passive sensing by estimating the point/extended target angle relative to the IRS using echo signals obtained via the BS-IRS-target-IRS-BS link. In addition, the IRS multi-view sensing was introduced in \cite{9966507} to further improve target detection and estimation, while the authors in \cite{9732186} demonstrated the feasibility of using IRS for networked target localization. Furthermore, the authors in \cite{9133157} optimized the configuration of IRS elements to create multiple independent paths, thereby providing richer information for enhanced human posture recognition. IRS‑aided imaging algorithms for human gesture and posture recognition have been studied in \cite{li2019machine, usman2022intelligent}. Moreover, the authors in \cite{10618967} proposed a device-free sensing system that integrates additional radio links provided by IRS into conventional radio tomographic imaging systems, thereby reducing the number of measurement nodes required for accurate image reconstruction.
High‑frequency sensing exploits the wide bandwidth and large array apertures available at mmWave and THz bands to achieve fine spatial and temporal resolution \cite{8240645}. 
Indoor scenarios often receive attention due to the limited transmission range at these frequencies. Although global navigation satellite systems (GNSS) provide acceptable outdoor localization, their performance degrades indoors. In this context, IRS enables precise localization and alleviates congestion caused by obstructions. For example, the IRS was used to construct a radio map of the propagation environment to detect the locations of both passive and active indoor users in \cite{10095766}. Similarly, the authors in \cite{9456027} and \cite{9427098} employed IRS to create favorable measurement conditions that improve the accuracy of indoor location estimation using received signal strength (RSS)-based techniques. In addition, the spherical wavefront propagation in the NF of THz systems was studied with the assistance of a passive IRS. 

Despite its advantages, the performance of IRS passive sensing is practically limited by the severe path loss inherent in the cascaded echo link. To overcome this limitation, IRS semi-passive sensing was proposed in \cite{shao2022target} (see Fig. \ref{environment}(b)). In this approach, additional low-cost active sensors are installed on the IRS so that it can both reflect BS signals and directly receive echo signals from the target. This dual functionality significantly reduces propagation path loss compared with conventional IRS passive sensing, allowing target detection based on echo signals received along the BS-IRS-target-IRS link. Recent IRS semi-passive sensing works can be
roughly divided into the two main categories, namely, sensing/reflection algorithm design and sensing performance analysis.
In the first category, the authors in \cite{10284917} examined target localization using echo signals from the BS-IRS-target-IRS link by optimizing the passive reflection matrix to maximize received power and proposed a low‑complexity location estimation algorithm. In \cite{10621068}, the authors focused on multi‑target direction of arrival (DoA) estimation in a semi‑passive IRS‑assisted sensing system. Rather than relying only on IRS reflecting elements for DoA estimation as in previous studies, they exploited information from both passive and active elements via an atomic norm minimization scheme. In the second category, the authors in \cite{song2023fully} evaluated SNR performance for target detection and Cramér-Rao bound (CRB) for DoA estimation, and demonstrated that fully passive IRS sensing outperforms the semi‑passive scheme when the number of IRS passive elements exceeds a specified threshold. Moreover, it is demonstrated in \cite{qiaoyan} that for extended targets, the CRB of the semi-passive IRS system is lower than that of the fully passive IRS if \(K < \mathrm{tr}\bigl((\mathbf{G}_r^H\mathbf{G}_r)^{-1}\bigr)\), with $K$ being the number of sensors and $\mathbf{G}_r$ denoting the IRS-BS channel.

Nevertheless, when the BS lies far from the IRS, IRS semi-passive sensing suffers from reduced accuracy. To overcome this limitation, IRS self-sensing was proposed in \cite{shao2022target} (see Fig. \ref{environment}(c)). In this architecture, the IRS controller extends its reflection control and information exchange functions to include transmission of sensing probes, while dedicated sensors on the surface capture target echo signals. Consequently, the IRS performs target localization independently, without relying on any BS for signal transmission or reception. Although IRS self-sensing incurs higher hardware and energy costs than IRS passive or semi-passive sensing, it offers improved performance because propagation paths are shorter. This approach has recently received considerable attention \cite{shao2022target,yumeng,10504753,10780951}. In particular, the authors in \cite{shao2022target} applied an IRS self-sensing to DoA estimation. They optimized IRS phase shifts to maximize received power at the IRS sensors and thereby minimize mean square error (MSE) for angle estimation. Then, the authors in \cite{yumeng} extended the IRS self-sensing structure from one IRS surface to a multi-surface case, where multiple active source controllers are installed, with one active source controller in each IRS surface. The IRS elements at each surface are independently illuminated by their aligned active source controller and reflect the impinging signals for radiation, thereby achieving full-space coverage for sensing targets distributed in three‑dimensional (3D) space.
Next, the joint problem of localization and channel estimation was considered in \cite{10504753} for a self-sensing IRS‑aided mmWave ISAC system, where a novel angle‑based sensing turbo variational Bayesian inference algorithm was designed to simultaneously obtain the approximate marginal posteriors of the target sensing and communication channels.
Furthermore, inspired by IRS self-sensing in \cite{shao2022target}, the authors in \cite{10780951} proposed a multifunctional IRS architecture with simultaneous self-sensing, reflection, refraction, and amplification functions. They investigated a signal-to-interference-plus-noise-ratio (SINR) maximization problem by optimizing IRS reflection coefficients and showed that under the same total power budget, the multifunctional IRS with self-sensing function attains 52.2\%, 73.5\%, and 60.86\% sensing SINR gains over active IRS (see Fig. \ref{advanced}(b)), passive IRS, and simultaneously transmitting and reflecting IRS (STAR-IRS) (see Fig. \ref{advanced}(d)), respectively.

The above works on IRS-enabled sensing have primarily concentrated on scenarios with a single IRS, which is limited in coverage by round-trip path loss. In real-world deployments, a BS may encounter several sensing coverage gaps because of distributed obstructions. To achieve full spatial sensing, various target localization and beamforming design methods have been proposed for systems where multiple IRSs cooperate to extend coverage over different LoS blocked regions (see Fig. \ref{environment}(d)). These investigations can be broadly categorized into two groups, namely multiple passive IRSs sensing and multiple semi-passive IRSs sensing. The first category addresses the use of multiple passive IRSs for sensing. For example, the authors in \cite{9963716} presented a multi-IRS-aided radar system aimed at improving the estimation accuracy for a moving target, where Doppler aware IRS phase shifts are obtained by minimizing the scalar
A-optimality measure of the joint parameter CRB matrix. The results showed that deploying multiple IRSs with these optimized phase shifts yields higher target sensing accuracy than non-IRS and single-IRS configurations. In the second category, each semi‑passive IRS is equipped with dedicated sensors to estimate the target’s DoA relative to itself. For instance, the authors in \cite{zihengmul} proposed a collaborative localization system that employs multiple semi-passive IRSs to locate one or more targets via time of arrival measurements by exploiting the geometric relationship between time delay and position. Their findings revealed that even a small number of semi-passive IRSs can produce effective multi-angle observations in space. Furthermore, a multi-semi-passive IRS time division sensing framework was proposed in \cite{xujiecrb} to schedule each semi-passive IRS to operate during separate time slots, thereby preventing overlapping sensing intervals and mitigating cross-interference between different IRSs. However, coordinating signals and control across multiple IRSs adds system complexity, especially in heterogeneous network environments, which can increase latency and reduce efficiency. Moreover, signal synchronization among IRSs poses an additional hurdle, as it demands precise timing to avoid interference and to maintain localization accuracy. These challenges remain open problems for future research.

\begin{figure}[t!]
	\centering
	\includegraphics[width=0.35\textwidth]{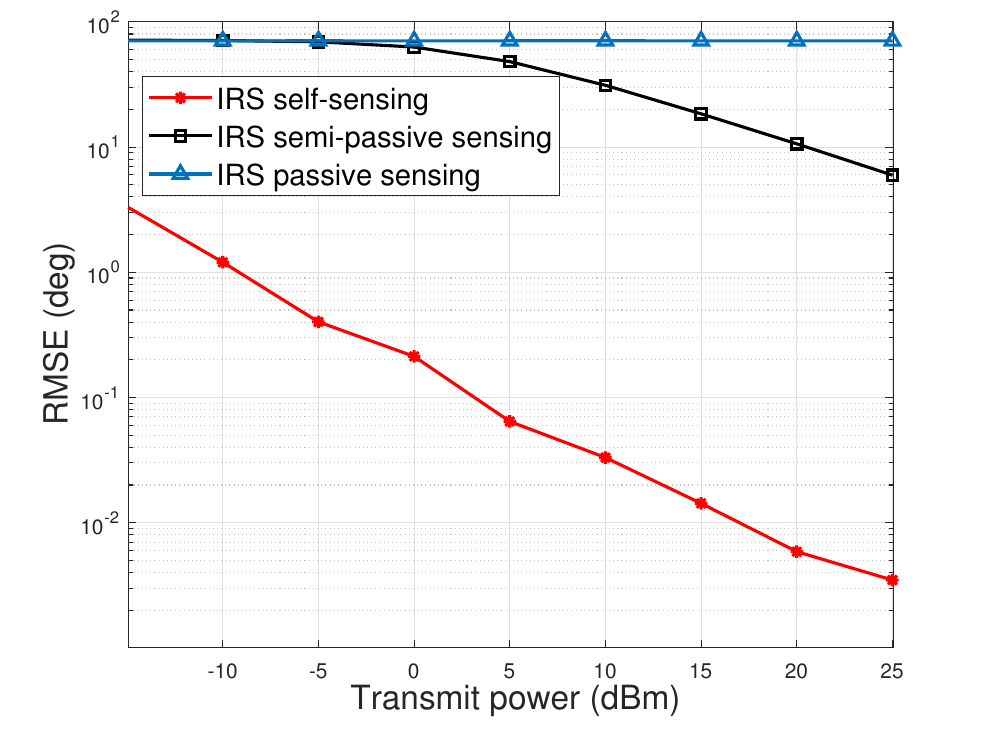}
    \vspace{-5pt}
	\caption{RMSE versus transmit power.}
	\label{semicom}
    \vspace{-10pt}
\end{figure}

To compare the performance of IRS passive sensing, IRS semi-passive sensing, and IRS self-sensing systems, Fig. \ref{semicom} compares the root mean square error (RMSE) performance of the considered systems when estimating target DoA using the multiple signal classification (MUSIC) algorithm within a specified Doppler bin \cite{shao2022target}. Two targets are positioned at azimuth angles of $60^\circ$ and $65^\circ$. The figure demonstrates that the IRS self-sensing architecture achieves substantially lower RMSE across the entire transmit-power range. IRS semi-passive sensing delivers moderate accuracy at high transmit powers but degrades as power decreases. In contrast, IRS passive sensing cannot resolve DoA in a pure LoS channel because it relies on two distinct propagation paths along the BS-IRS-target link.

\subsubsection{Target-mounted IRS}
\begin{figure*}[!t]
	\centering
	\includegraphics[width=0.7\textwidth]{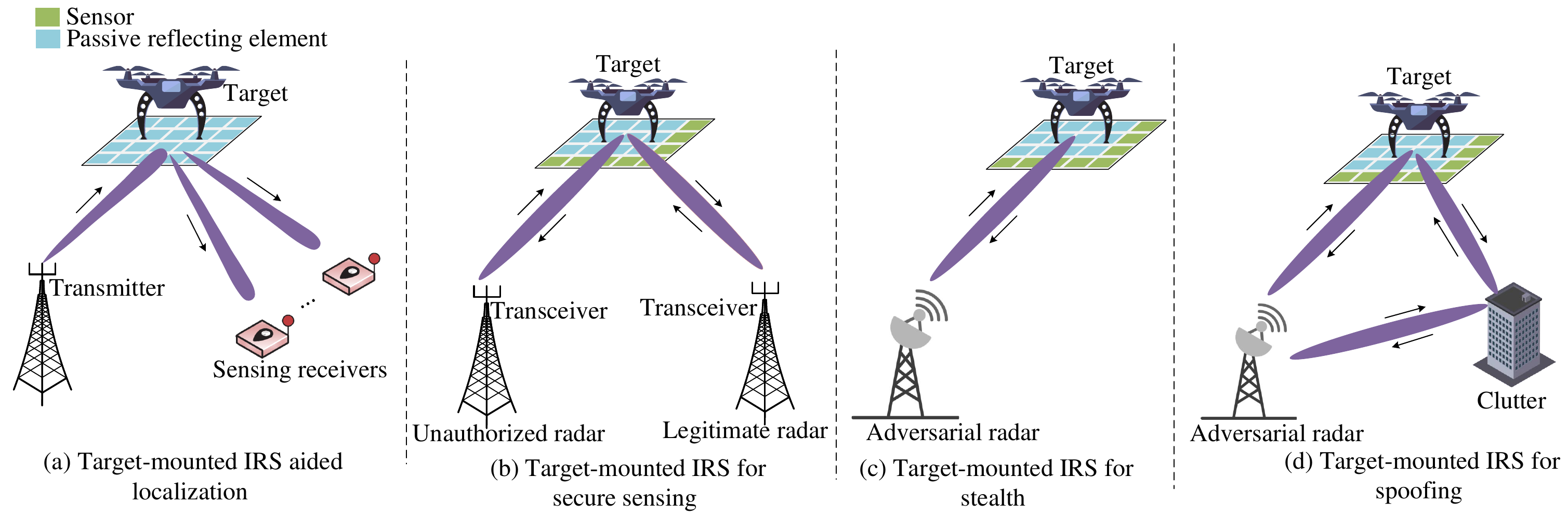}
	\caption{Target-mounted IRS for sensing.}
	\label{target_mounted}
\end{figure*}

For prior IRS-aided wireless sensing systems, the IRS is treated as an anchor node, and sensing performance depends largely on direct and IRS-reflected echo signals from the target to the receiver. This approach may be weak in practice because of the generally small RCS of the target and its random signal reflection. Recently, the potential of a target-mounted IRS, where the IRS is installed on the sensing target (see Fig. \ref{target_mounted}), has been revealed in IRS-aided sensing systems \cite{shao2024target, mountlocal, shao2023enhancing}. Target-mounted IRS offers enhanced sensing by using its controllable signal reflection and high spatial resolution provided by a large aperture. For example, in \cite{mountlocal}, a fully passive IRS is mounted on the target, and its location and orientation are estimated via a tensor algorithm. Owing to the improved target RCS and controlled reflection, this method significantly enhances sensing performance even with few sensing receivers. Moreover, for extended targets, accurate position and orientation determination becomes possible when multiple IRSs are strategically placed on target surfaces \cite{kaitarget}. In addition, the phase shift control of target-mounted IRS allows for modulating information on echo signals to achieve identifier correlation for multiple targets, which is useful for target association in dense scenarios.

Target-mounted IRS has been proposed to improve wireless sensing security. As shown in Fig. \ref{target_mounted}(b), secure sensing is aimed at enhancing the performance of the legitimate radar station (LRS) while preventing detection by unauthorized radar stations (URS). To this end, the authors in \cite{shao2024target} proposed a secure sensing protocol using a target-mounted semi-passive IRS. In this protocol, each coherent processing interval is divided into two steps. In the first step, IRS sensors estimate the LRS and URS channels and waveform parameters with all IRS elements turned off. In the second step, based on the estimated parameters, the IRS reflection is designed to boost the received signal at the LRS receiver while reducing that at the URS receiver. This design provides significant gain at the LRS and notable suppression at the URS compared with scenarios without a target-mounted IRS. Furthermore, the role of target-mounted IRS in both sensing enhancement and suppression was investigated in \cite{kong2024signal} by jointly optimizing the transmit beamforming and the IRS reflection coefficients. Note that traditional methods employing retrodirective arrays \cite{ana} and absorbing material \cite{ret} do not afford the same flexibility in controlling the reflected signal direction as the target-mounted IRS.

In addition to enhancing sensing accuracy and security, target-mounted IRS can contribute to target stealth (see Fig. \ref{target_mounted}(c)) and radar detection spoofing (see Fig. \ref{target_mounted}(d)). The basic idea behind IRS-aided stealth and spoofing is to manipulate IRS reflection to neutralize or disguise the target’s echo signals that return to adversarial radars, thereby lowering the likelihood of accurate detection \cite{spoof,sec1,xu2025intelligent}. By properly adjusting the amplitudes and phase shifts at the IRS, the echoes from both the IRS and the target surface can be combined destructively to cancel the effective signal toward adversarial radar, effectively reducing the target’s detectability and achieving stealth \cite{xu2025intelligent}. Alternatively, the IRS can be configured to eliminate the target’s echo signals in the direction of the adversarial radar while directing reflected signals toward clutter to form decoy targets, thus increasing the possibility of false detection and affecting spoofing \cite{spoof}. When multiple distributed radars are present, the IRS reflection can be designed as an angle-selective reflection pattern that suppresses several probing signals in different directions. Compared with conventional stealth and spoofing techniques, the low-profile, lightweight, and conformal geometry of a target-mounted IRS provides the advantage of dynamically adjusting reflection properties, making it an effective solution in time-varying and highly dynamic detection environments caused by high-mobility targets and rapidly changing adversarial radar modes.

\subsubsection{Advanced IRS}
In addition to the IRS sensing architectures discussed above, several new designs have been proposed to overcome their limitations and enhance performance, including beyond diagonal IRS (BD-IRS) sensing, active IRS sensing, holographic IRS sensing, and STAR-IRS, as follows:
\begin{figure*}[!t]
	\centering
	\includegraphics[width=0.7\textwidth]{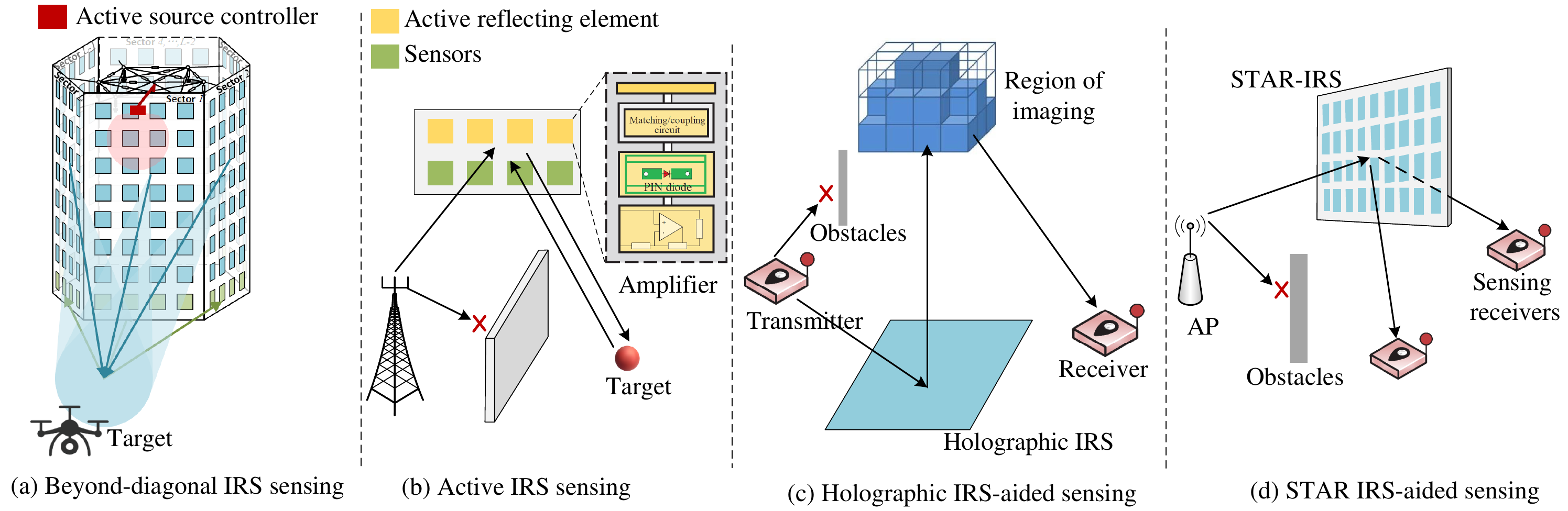}
	\caption{Advanced IRS for sensing.}
	\label{advanced}
    \vspace{-10pt}
\end{figure*}

\begin{figure}[!t]
	\centering
	\includegraphics[width=0.35\textwidth]{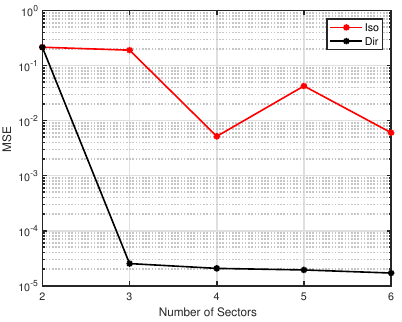}
	\caption{The MSE of BD-IRS sensing versus the number of sectors.}
	\label{bdirs}
    \vspace{-10pt}
\end{figure}

Most existing IRS sensing studies concentrate on using a simple IRS model with a diagonal phase shift matrix, where each IRS element is connected to its own reconfigurable impedance without any inter-element connections. Such a lossless IRS only supports the reflection of signals toward the same side, which limits the sensing coverage. To overcome this limitation and to enhance the performance gain of IRS sensing, BD-IRS self-sensing (see Fig. \ref{advanced} (a)) was proposed in \cite{yumeng}, where an active source controller is installed in one sector to probe signals toward that sector. In addition, the IRS elements across the $L$ sectors (with $L \leq 2$) are connected through a reconfigurable impedance network.  Multiple sectors enable the probing signal to be simultaneously reflected by each sector and scattered by the others, thereby facilitating full‑space radiation. \cite{yumeng}. With BD-IRS sensing, a BS can achieve full spatial sensing coverage even in the presence of distributed obstructions. As illustrated in Fig. \ref{bdirs}, the multi-sector BD-IRS self-sensing system, particularly the 4-sector architecture, achieves full-space sensing capability beyond that of single-sector or double-sector IRS configurations. Moreover, with an isotropic antenna pattern, the overall MSE for odd $L$ is worse than that for even $L$ in the multi-sector BD-IRS self-sensing system. This occurs because the multi-sector BD-IRS geometry with an even $L$ benefits from a relatively more uniform MSE as a function of the target DoA. Furthermore, when a directive antenna pattern is employed, the overall MSE consistently improves as $L$ increases. This improvement results from the higher directivity provided by directive antenna patterns with larger $L$.

As discussed earlier, the reflected sensing signal by a conventional passive IRS with unit modulus constraint suffers from severe product-distance path loss for sensing. In other words, the cascaded path loss in the transmitter-IRS-receiver link is the product of those in the transmitter-IRS and IRS-receiver links, respectively. Although equipping the IRS with massive reflecting elements or placing it close to the transceivers can alleviate this problem, these measures impose significant challenges in practical IRS deployment and result in higher sensing overhead and complexity. To tackle this issue, the active IRS has been proposed in \cite{9377648,9998527,10417102} to enable simultaneous signal reflection and amplification. As shown in Fig. \ref{advanced}(b), active IRS comprises a number of meta-atoms equipped with reflection-type amplifiers. Compared with conventional passive IRS without signal amplification, active IRS can alter the signal's phase and amplify its amplitude to compensate for the severe cascaded path loss for sensing with moderately higher power consumption and hardware cost. In addition, the amplitude control with power amplification provided by active IRS offers additional DoF to achieve finer-grained reflective beamforming for wireless sensing relative to passive IRS. These appealing features of active IRS hold great potential to significantly enhance the performance of wireless sensing applications, such as indoor localization, target detection, and imaging. For example, the authors in \cite{song2024active} studied target detection using an active IRS to determine target presence, and an optimal detector was derived under the Neyman-Pearson theorem by combining the BS transmit signal with IRS reflection noise. Their analysis confirms that amplification at the active IRS and utilization of reflection noise in detector construction substantially enhance detection probability (DP).

In the above works, the authors modeled the IRS as being composed of a series of discrete metamaterial elements, each applying a phase shift to the incident signal. However, this approach is overly simplistic and does not adequately capture the interaction mechanism between the IRS and the incident electromagnetic field, nor does it fully analyze the electromagnetic characteristics of the IRS. This mismatch can result in sensing algorithms that inadequately capture the IRS’s true behavior and produce inaccurate performance evaluations. To overcome this limitation, a holographic IRS has been proposed in \cite{9374451,10606489,10719631}, where the IRS is modeled as a physically continuous surface with a continuously varying phase shift pattern, which can be approximated as a finite space with a massive number of elements. The holographic IRS sensing is capable of reshaping the incident electromagnetic waves toward the desired target direction and provides a reliable and universal electromagnetic model that accurately reflects the IRS's electromagnetic nature (see Fig. \ref{advanced}(c)). Compared with conventional IRS sensing, the main advantages of holographic IRS sensing lie in its greater control and flexibility when designing the scattered field properties, thereby achieving higher sensing accuracy. In addition, this model offers a new perspective for analyzing the electromagnetic characteristics of IRS sensing based on Maxwell equations and electromagnetic theory. Consequently, holographic IRS sensing has attracted significant research interest. For instance, the authors in \cite{gan2022near} investigated a holographic IRS-assisted mmWave near field (NF) localization system and showed that the localization accuracy increases quadratically with the holographic IRS size. Then, the authors \cite{sun2024computational} designed a holographic IRS for computational imaging by dynamically altering the IRS phase shifts to focus the scattered fields onto different target segments for facilitating robust imaging performance. 

Conventional IRS can only reflect the incident signal and cannot transmit it. As a result, when the radar and target are located on opposite sides of the IRS, a single IRS cannot achieve full target sensing. To overcome this limitation, STAR-IRS sensing (see Fig. \ref{advanced}(d)) has been proposed in \cite{he2022simultaneous}. Unlike traditional passive IRS sensing, each STAR-IRS element is capable of both refracting and reflecting the incident probing signal, which eliminates the need to restrict deployment to specific geographical areas and enables full-space target coverage \cite{he2022simultaneous,li2024multipath,li2023csinet}. Consequently, STAR-IRS can enhance the localization and sensing capabilities of wireless networks, particularly in confined spaces such as buildings, by overcoming signal barriers and providing comprehensive spatial coverage. In addition, STAR-IRS can improve mobile robot localization and data transfer speeds in intelligent factory environments.

\subsection{Sensing Performance Metrics}






An IRS sensing system is designed based on several performance metrics that include accuracy, robustness, and resolution, and so on. 

First, accuracy refers to metrics that evaluate how closely an estimated target location matches the ground truth and the precision in determining that location, often represented by the standard deviation or confidence interval. The deployment geometry of the IRS and the relative position of targets with respect to BSs influence accuracy alongside link-level SNR. Multiple accuracy performance metrics can be optimized to enhance positioning performance, such as the received SNR and the CRB on target location estimation. The authors in \cite{song2023fully} investigated the sensing performance of IRS-enabled NLoS sensing systems using passive and semi-passive IRS. 
Specifically, the sensing SNR scales significantly with the number of reflecting elements $N$. In contrast, the CRB decreases inversely with the number of elements. Both metrics benefit from passive beamforming; however, while the SNR emphasizes power enhancement, the CRB reflects the impact of phase differences across the array on estimation precision.

Second, resolution denotes the separability of completely correlated radio propagation channels in at least one domain. When signal paths cannot be resolved, they merge into a single path, which limits accuracy regardless of SNR and yields performance below analytical predictions. Resolution is determined by physical resource constraints. Angle resolution depends on the antenna array aperture, while delay or distance resolution is governed by the available bandwidth \cite{9893187}. High resolution does not preclude ambiguity or nonidentifiability in localization \cite{10044963}. Ambiguity arises when multiple distinct locations produce similar propagation characteristics and is usually resolved with prior knowledge or external signals. Nonidentifiability poses a greater challenge because it yields many equally valid solutions that external information alone cannot eliminate.
To quantify these effects in IRS passive sensing, a comprehensive resolution analysis in the context of least‑squares solutions was conducted in \cite{wang2024toward} via condition‑number analysis of the sensing operators.

Third, the robustness of a localization algorithm can be evaluated by its performance under environmental conditions such as multipath fading, interference, and mobility. Robustness metrics include availability, latency, and update rate \cite{9782674}. It is also crucial to compare the energy consumption of IRS-assisted localization techniques to systems without IRS. Energy-efficient strategies aim to reduce power usage while maintaining localization accuracy. Finally, computational complexity, measured by processing time, memory requirements, and algorithm workload, directly influences latency and update rate. Lower complexity reduces processing delays and enables more frequent updates, helping the system maintain high availability and accuracy as environmental conditions change.

\subsection{Detection and Estimation Algorithms}
Sensing fundamentally involves two tasks, namely, target detection and parameter estimation. Specifically, target detection involves making statistical decisions regarding the presence or classification of sensed objects, ranging from binary detection, e.g., target present/absent, to multi-class categorization, e.g., weight classification as light, medium, or heavy \cite{ZhangJAsurvey}. The corresponding performance is measured by DP and false alarm probability. Parameter estimation extracts physical target characteristics from the received echo signals, including distance, velocity, angle, position, and temperature, with performance metrics such as MSE or the CRB. The two processes exhibit strong interdependence, collectively determining overall system sensing capability.  Target detection serves as a preprocessing stage, identifying regions of interest for subsequent parameter analysis, which effectively reduces the computational complexity of parameter estimation.  Conversely, the accuracy of parameter estimation directly improves the quality of detection. First, accurate parameter extraction can distinguish true targets from false alarms. Second, precise parameter estimation helps adaptive threshold optimization in detection algorithms to balance DP and false alarm rate under different environmental conditions. 

IRS offers substantial benefits for sensing systems, particularly in target detection and parameter estimation. In \cite{YuTSP}, the proposed framework implements a hierarchical two-stage detection and estimation process. Initial target detection and angle of departure (AoD) estimation are achieved through codebook-based 3D IRS beamforming with delay spectrum peak-based beam training, followed by precise range and velocity estimation via adaptive beam refinement.

\subsubsection{Detection} IRS can dynamically generate directional beams with high gain by precisely controlling the electromagnetic characteristics of its elements. It allows for accurate illumination of target regions, significantly improving detection sensitivity and range resolution by concentrating radar energy in desired directions. In the following, we discuss three sensing architectures for target detection, namely passive, semi-passive, and active IRS.

\paragraph{Target detection with passive IRS} Target detection was investigated for radar systems assisted by a single IRS \cite{detection3,9732186} and by multiple IRSs \cite{detection1,detection2}. In \cite{detection3}, the interference was suppressed in cluttered environments via a joint optimization of the transmit waveform, the receive filter, and IRS beamforming to maximize the SINR. The work \cite{9732186} investigated the effect of IRS deployment on a collocated multiple-input multiple-output (MIMO) sensing system and showed that locating the IRS in the NF of the radar transmitter and receiver reduces indirect link path loss, while FF placement yields only marginal gains for targets within the radar main lobe. The same study demonstrated that, in a monostatic configuration, a single IRS can serve both forward and backward reflection with identical phase settings, simplifying hardware design and control. To overcome LoS blockage, multiple IRSs were deployed in \cite{detection1} and \cite{detection2}, resulting in superior detection performance over a single IRS setup. In \cite{detection1}, IRSs positioned close to the radar transmitter and receiver steered energy toward the target and collected echo signals. Orthogonal transmit waveforms and the generalized likelihood ratio test (GLRT) guided the selection of reflection coefficients to maximize received echo SNR and improve DP.
In \cite{detection2}, the authors investigated a multi-IRS-aided distributed MIMO system, maximizing target DP by optimizing the IRS passive beamforming under a fixed false alarm probability constraint.
The works discussed above rely on the receiver to passively analyze and separate characteristics of the echo signals, including the directionality, delay, and Doppler. These methods rely on the complexity of receiving algorithms and flexibly respond to unknown targets, but they impose high demands on hardware and computing power. A novel signal preprocessing strategy was adopted at the transmitter in \cite{9938373} to improve detection performance by establishing a mapping between signature sequences and multi-target directions. It benefits from actively designing signals and reducing the computational burden on the receiver, making it suitable for multi-target or dynamic target detection scenarios. 

The deployment of multiple passive IRSs introduces a critical challenge in sensing systems, e.g., the BS receives desired target echoes from monitored regions, and it simultaneously captures interference signals from non-target areas through unintended IRS reflections. This cross-region interference substantially degrades detection reliability by introducing false positives and obscuring weak targets. To overcome this limitation, advanced interference-aware detection algorithms can enhance discrimination capabilities in spectrally congested environments, while beamforming optimization techniques suppress interference through spatial filtering. A practical alternative employs time-division sensing protocols that sequentially activate IRSs across distinct regions, effectively isolating interference through temporal separation. However, these approaches face fundamental performance-complexity trade-offs. Recent implementations demonstrate superior performance through semi-passive or active IRS architectures, where integrated sensors enable localized echo reception at each surface while physical obstructions between regions naturally attenuate cross-interference.

\paragraph{Target detection with semi-passive IRS} The DP was maximized in \cite{o3} by joint beamforming design, subject to the constraints on the BS’s transmit signal and the IRS’s reflection. Under the aforementioned LoS channel and point target models, the corresponding joint optimization is equivalent to that for SNR maximization. Specifically, the optimal transmit beamforming follows the maximum-ratio transmission (MRT) technique and the optimal reflective beamforming aligns the phase of signals toward the target. Moreover, critical insight reveals a consistent positive correlation between DP and the number of IRS reflecting elements.
	
\paragraph{Target detection with active IRS} Due to its signal amplification capability, the active IRS delivers superior detection performance compared to its fully passive and semi-passive counterparts. In \cite{song2024active}, the authors investigate a target detection system in which an active IRS equipped with reflecting elements and sensors processes echo signals over the BS-IRS-target-IRS link to determine the presence of the target. The results demonstrate that active IRS amplification, incorporation of reflection noise in detector design, and joint beamforming optimization markedly improve DP. Unlike the sensing SNR method, which uses only the BS transmit signal and treats reflection noise as interference, the detector in \cite{song2024active} exploits both the BS transmit signal and the active IRS reflection noise. 

\subsubsection{Parameter estimation} In parameter estimation, IRS contributes by optimizing electromagnetic wave propagation paths to mitigate multipath effects and interference. It enhances measurement stability and reliability while facilitating more accurate extraction of target parameters. Furthermore, the IRS can be integrated with advanced signal processing algorithms to refine radar echo signals to achieve superior estimation precision. The following analysis starts with parameter estimation for the single-target case and then extends to the multi-target scenario.

\paragraph{Single-target localization} To address the fake target angle issue caused by IRS reflections, the work \cite{locsensing0} proposed a sparse Bayesian learning method in a passive IRS-aided sensing system by leveraging angular reciprocity between downlink and uplink channels. The fake and true angles are then distinguished by the common and independent sparse elements recovered via the expectation-maximization (EM)-based algorithm, respectively.
In a passive IRS-aided indoor localization system, two key practical challenges should be considered: namely, transmitter position inaccuracy from distance measurement bias and unknown IRS phase offsets. To solve these problems, the developed calibration technique in \cite{locmobil} exploits the linear relationship between measurement bias and localization error, while a novel signal processing method isolates and removes non-IRS components, enabling robust indoor positioning. To further enhance angle and range estimation accuracy, previous studies \cite{10284917, semilocal} proposed a semi-passive IRS architecture to mitigate path loss effects. Moreover, the corresponding system parameters, such as the number of IRS passive elements and sensors should be carefully designed under given budget constraints. The advantages of active IRS for target localization were highlighted in \cite{actlocal}, which proposed a neural network-based data-driven approach using short pilot sequences. The proposed method first probes with broad beams, then adaptively refines focus onto desired low-dimensional channel components for precise angle-of-arrival estimation. The work was extended to an indoor environment case in \cite{LiCL}, where multipath information was captured to further improve localization accuracy.

\paragraph{Multi-target localization} Single-target research can be extended to multi-target setups in complex fading environments with additional signal information and introducing novel signaling protocols in FF, NF, and mixed NF and FF scenarios \cite{multitarget,multitarget2,10149471,MixedField,nearfar}. Existing research demonstrates the effectiveness of a single IRS by creating virtual LoS paths for improving FF multi-target localization. Although multiple IRSs can further improve coverage and measurement accuracy for multi-target sensing, they introduce challenges in distinguishing signals reflected from different targets. To address these issues, a heuristic-based matching algorithm was provided in \cite{multitarget} to uniquely identify target locations from potential candidates, where the IRS forms a passive directional beam to scan the space to locate targets.  To associate multiple locations accurately, the work \cite{multitarget2} solved a non-convex atomic norm minimization (NC-ANM) problem by exploiting target sparsity in the spatial domain for DoA estimation and multi-target localization. 
To overcome resolution limitations in dense multi-target scenarios, researchers have also focused on IRS-assisted sensing systems based on high-frequency bands. For NF localization, a localization method based on the MUSIC algorithm was proposed in \cite{10149471} for an ultra-large IRS-aided THz system to estimate angles of arrival (AoAs) and distances of users located in the Fresnel regions of the IRS and BS. It significantly reduces the need for distributed arrays and their complex synchronization. In addition, the NF beam-focusing effect can be exploited to improve the sensing SNR, enabling more accurate parameter estimation.
In practical applications such as medical imaging, acoustic sensing, seismic exploration, and electronic supervision, mixed NF and FF sources may occur or sources may exhibit both NF and FF characteristics \cite{MixedField}. Although conventional techniques prove inadequate for such scenarios due to their inherent limitation to pure FF or NF resolution, recent work \cite{nearfar} addresses this through IRS-aided localization. The proposed three-phase localization protocol distinguishes NF/FF targets by leveraging the BS-received signals and localizes these targets while balancing aperture loss, computational complexity, and estimation accuracy. 

\subsubsection{Target tracking} 
For tracking tasks, the IRS phase shifts are required to be updated based on the targets' positions relative to the IRS \cite{tracking2,tracking,tracking1}. In \cite{tracking2}, an effective tracking technique is proposed that uses the phase differences between IRS subpanels to estimate azimuth and elevation AoDs. By using exclusively separated subpanels, as subsets of reflecting elements used to steer signals toward specific directions, the IRS enables simultaneous multi-directional reflection when independently controlled. However, traditional tracking algorithms exhibit limitations in complex environments, particularly when handling nonlinear trajectories in indoor scenarios affected by multi-path propagation and NF effects, leading to reduced localization accuracy. Furthermore, these methods inadequately leverage temporal correlation in measurement data, particularly in recognizing long-term dependencies, which restricts prediction accuracy and model adaptability. Recent studies have employed neural network architectures for scenarios with available movement pattern data. As shown in \cite{tracking}, IRS information can be processed through a data-driven framework combining a convolutional neural network (CNN)-based information reconstruction with integrated feature extraction and mobile tracking modules. Despite this, the requirement for training data limits the applicability of such models, especially when user movement patterns change over time. To address these issues, alternative approaches like the codebook-based tracking algorithm in \cite{tracking1} offer generalized solutions for nonlinear movement patterns. This method performs iterative maximum likelihood (ML)-based direction estimation at fixed intervals, enabling adaptive reconfiguration of IRS phase shifts without requiring extensive training data.

\begin{table*}[!t]
\centering
\renewcommand{\arraystretch}{1.35}
\small
\caption{Comparison of Sensing Algorithms and Beamforming Designs for IRS‑aided Sensing Systems}
\label{tab:SensingComparison}
\resizebox{\linewidth}{!}{
\begin{tabular}{|
    p{0.1\linewidth}|
    p{0.08\linewidth}|
    p{0.07\linewidth}|
    p{0.075\linewidth}|
    p{0.11\linewidth}|
    p{0.08\linewidth}|
    p{0.17\linewidth}|
    p{0.19\linewidth}|}
\hline
\textbf{Application} & \textbf{Reference} & \makecell{\textbf{Link}\\\textbf{Status}} & \textbf{Scenario} & \textbf{IRS Type} & \makecell{\textbf{Number}\\\textbf{of Targets}} & \textbf{Design Objectives} & \textbf{Algorithms} \\
\hline

\multirow{6}{=}{\parbox[t]{\linewidth}{\raggedright Target detection}}
& \cite{9732186} & LoS, NLoS & Far/near field & multi-passive IRSs & Single
& Maximize DP under fixed false alarm
& Alternating maximization \\ 
\cline{2-8}
& \cite{YuTSP} & LoS only & Far-field & Single passive IRS & Single
& Maximize beam gain along direction
& Beam-refinement \\ 
\cline{2-8}
& \cite{detection3} & LoS, NLoS & Far-field & Single passive IRS & Single
& Maximize SINR
& MM, FPP, KKT \\ 
\cline{2-8}
& \cite{detection1} & LoS, NLoS & Far-field & multi-passive IRSs & Single
& Maximize DP
& Projection + PMLI \\ 
\cline{2-8}
& \cite{o3} & LoS, NLoS & Far-field & Passive/semi/ active IRS & Single
& Compare performance across architectures
& MLE, GLRT, SDR, SCA \\ 
\cline{2-8}
& \cite{9938373} & LoS, NLoS & Far-field & Single passive IRS & Multi
& Maximize min beam-pattern gain
& Penalty-based algorithm \\
\hline

\multirow{7}{=}{\parbox[t]{\linewidth}{\raggedright Parameter estimation}}
& \cite{shao2022target,shao2024target} & LoS only & Far-field & Semi-passive IRS & Single
& Improve MSE of sensing
& MUSIC \\ 
\cline{2-8}
& \cite{locmobil} & LoS, NLoS & Far-field & Semi-passive IRS & Single
& Optimize prototype + signal power
& Calibration via linear bias error model \\ 
\cline{2-8}
& \cite{multitarget2} & LoS only & Far-field & multi-passive IRSs & Multi
& Improve RMSE of sensing
& NC-ANM \\
\cline{2-8}
& \cite{10149471} & LoS only & Near-field & Single passive IRS & Multi
& Minimize cost function
& LS \\ 
\cline{2-8}
& \cite{nearfar} & LoS only & Mixed near/far field & Single passive IRS & Multi
& Improve sensing coverage
& MUSIC \\
\hline

\multirow{3}{=}{\parbox[t]{\linewidth}{\raggedright Target tracking}}
& \cite{tracking2} & LoS, NLoS & Far-field & Single passive IRS & multi-mobile
& Maximize correlation
& ZC-based method \\ 
\cline{2-8}
& \cite{tracking} & LoS, NLoS & Near-field & Single passive IRS & Single mobile
& Minimize position error
& Deep learning \\ 
\cline{2-8}
& \cite{tracking1} & LoS only & Near-field & Single passive IRS & Single mobile
& Minimize direction error
& MUSIC, Kalman filter, codebook \\
\hline

\multirow{4}{=}{\parbox[t]{\linewidth}{Beamforming}}
& \cite{beamdesign} & LoS only & Far-field & Single passive IRS & Multi
& Maximize mutual info via space-time BF
& Sherman-Morrison-Woodbury, exhaustive search \\ 
\cline{2-8}
& \cite{coopopt} & LoS only & Far-field & multi-passive IRSs & Multi
& Minimize max CRB
& Polyblock algorithm \\ 
\cline{2-8}
& \cite{jointBFFeng} & LoS only & Near-field & multi-passive IRSs & Multi
& Minimize Tx power under localization constraint
& SDP, GP \\
\hline

\multicolumn{8}{|p{\textwidth}|}{
\footnotesize
\textbf{Abbreviations:} FPP: feasible point pursuit; KKT: Karush-Kuhn-Tucker;  
PMLI: power‑method‑like iterations; MLE: maximum likelihood estimation; 
NC‑ANM: non‑convex atomic-norm minimization; ZC: Zadoff-Chu;  
SDP: semidefinite programming; GP: geometric programming.
} \\
\hline
\end{tabular}}
\end{table*}

\subsection{Beamforming Design} 
The beamforming design of an IRS plays a crucial role in simultaneously improving the accuracy and resolution of target sensing \cite{shao2022target,beamdesign,coopopt,qiaoyan_sensor}. Since the reflection pattern of the IRS can be flexibly adjusted, it enables adaptive beamforming strategies according to the availability of prior target location information. When the target position is unknown, the IRS employs a quasi‑omnidirectional reflection mode, in which the phase shifts of its reflecting elements are randomly or uniformly distributed to radiate energy across the entire detection region. This approach, although reducing signal strength in any specific direction, provides full spatial coverage and is thus suitable for initial target discovery in unknown environments. Once coarse azimuth information becomes available, the IRS reconfigures its reflection phases to focus energy within specific angular ranges, forming directional wide beams that enhance the received signal quality around the target area. This adaptive transition from wide‑beam search to focused tracking enables accurate sensing while maintaining broad coverage. 

The intelligent reflective surface reconstructs the electromagnetic wave propagation path by dynamically adjusting its reflection coefficient, thereby realizing its passive beamforming in the spatial domain and consequently improving the angular resolution and multi-target detection capability of the sensing system. Furthermore, a larger IRS size can provide higher beamforming gain in the spatial domain. For instance, in \cite{coopopt}, an IRS mounted on the target is employed for localization, where a cooperative scheme that actively controls echo signals is introduced. The system minimizes the maximum CRB through joint optimization of target association, IRS phase shifts, and dwell time, resulting in enhanced sensing performance. The work \cite{shao2022target} investigated an IRS self-sensing configuration for target position estimation by optimizing the passive reflection matrix to maximize average received signal power. Beyond spatial-domain benefits, the IRS can also adjust the beamforming in the time domain to improve resolution. Specifically, the IRS dynamically switches its reflection to provide a wider range of beamforming vectors in the time domain. For example, the IRS adjusts its reflection direction in a time-sharing manner to enhance the continuity of target tracking. In the anti-interference scenario, the IRS can reduce the impact of interference on the sensing signal through time-domain reflection scheduling, such as avoiding periodic interference periods. By integrating space and time domain strategies, the joint design of IRS's spatial beamforming and dynamic scheduling over time can break through the performance limitations of a single domain, thereby achieving a comprehensive improvement in sensing capabilities. Building on this, \cite{beamdesign} proposes a beamforming framework that combines space domain beamforming for accuracy improvement and time domain beamforming to maximize the upper bound of the sensing mutual information. The results reveal that space-time beamforming increases the rank and eigenvalues of the equivalent sensing channel matrix, which enhances sensing resolution and accuracy.

Compared with optimizing only the IRS passive beam, recent studies have shown that jointly designing the IRS reflection and the BS transmit beam significantly improves performance \cite{detection3,beamfor1,jointBFFeng,jointBFLiu}. The joint approach enables precise beam alignment between the BS and the IRS to enhance signal power and provides richer spatial observation information, with greater advantages when the target position is unknown. However, practical problems remain, including maintaining LoS links between the BS and the IRS, ensuring accurate phase modulation, and sustaining tracking performance in dynamic, multipath environments. To tackle these challenges, \cite{detection3} proposed a joint optimization framework that simultaneously designs the space‑time transmit waveform, receive filter, and IRS reflection coefficients to maximize the minimum sensing SINR under power and configuration constraints. In \cite{beamfor1}, an alternating optimization (AO) method was used to iteratively update BS beamforming and IRS reflection, fully exploiting the spatial DoFs provided by multiple antennas. Furthermore, as demonstrated in \cite{beamfor1}, the IRS can enhance desired signal power and suppress interference at undesired locations. To facilitate implementation, the BS beam can remain fixed toward the IRS controller to maintain a stable link, while the IRS adaptively adjusts its phase shifts for beam scanning. This design offloads beam‑steering from the BS to the IRS, reducing computational complexity while preserving sufficient spatial DoFs for accurate target estimation.

\section{IRS-aided ISAC: Coexistence}
In conventional ISAC systems, the performance trade-off between communication and sensing functions presents a core challenge. Especially in Radar-Communication Coexistence (RCC) scenarios, system performance is severely constrained by mutual interference. IRS, by providing additional spatial DoFs, offers a new pathway to manage this trade-off and suppress interference. This section explores ISAC coexistence aided by IRS, including the fundamental ISAC performance trade-off, interference mitigation and optimization techniques in RCC, and the design of Dual-functional Radar-Communications (DFRC).

\subsection{ISAC Trade-off}
The performance trade-off between communication and sensing functions has been extensively studied in conventional ISAC systems. Existing research has established systematic theoretical frameworks to capture the inherent limitations in jointly optimizing sensing accuracy and communication throughput under shared resource constraints.

In non-IRS ISAC systems, mainstream approaches to trade-off modeling can be grouped into several lines of work. First, the CRB-rate framework has been widely adopted to characterize the Pareto boundary between localization accuracy and achievable rate. In both point-target estimation and multi-antenna MIMO settings, the CRB reflects the lower bound on sensing estimation error, while rate serves as the throughput metric for communication. Studies have revealed that beamwidth, array geometry, power allocation, and multi-target interference jointly affect the achievable balance between CRB and rate \cite{10217169,10251151}.

Second, another body of work focuses on non-parametric metrics such as probability of detection, bit error rate, and their coupling. Researchers have demonstrated how waveform design—particularly Orthogonal Frequency Division Multiplexing (OFDM) and constellation shaping—impacts the DP-rate or bit error rate (BER)-DP curves. Notably, the Kullback-Leibler divergence has been introduced as a sensing separability metric under varying modulation structures, offering robustness against noise and multi-target ambiguity \cite{10124135,10685511,10636778}.

Third, information-theoretic analyses have provided unified models of capacity-distortion trade-offs under Gaussian channels. These frameworks define joint capacity and estimation bounds under idealized channel conditions and prove that ISAC systems cannot simultaneously optimize both sensing and communication functions under finite bandwidth and power constraints \cite{10147248}. Furthermore, constellation-shaping-based waveform design has been explored to push the Pareto frontier outward within these bounds \cite{10685511}.

However, all of these models are derived assuming fixed wireless environments without IRS. The introduction of IRS significantly alters the system’s spatial DoFs and thus the nature of ISAC trade-offs. By precisely controlling the phase and direction of incident signals, IRS can enhance link reliability and sensing reflections without additional power or spectrum. This offers a new paradigm for potentially breaking through conventional ISAC performance bottlenecks.

While many existing IRS-ISAC studies focus on joint beamforming and optimization algorithms, few provide a systematic understanding of how IRS alters the CRB-rate or DP-rate boundary itself. For example, the IRS can amplify echo signal strength or suppress cross-interference between radar and communication beams. However, without unified performance metrics, it remains unclear whether these enhancements improve the overall trade-off or merely benefit one function at the expense of the other. Moreover, current literature often oversimplifies IRS-specific parameters—such as the number of elements, phase quantization levels, and deployment geometry—without quantifying their effects on the trade-off frontier.

Unlike traditional ISAC systems, where performance is constrained by a static channel, IRS enables dynamic manipulation of spatial energy distribution, potentially reshaping the balance between communication and sensing objectives. For instance, IRS beam focusing may boost DP while preserving data rate, yet the extent of such improvement remains theoretically undefined. To this end, future IRS-ISAC research should aim to develop trade-off modeling frameworks that incorporate IRS configurations as core variables, extending CRB-rate and DP-rate models to IRS-aware formulations.

Such extensions may reveal whether the IRS can shift the Pareto boundary, enabling ISAC systems to achieve previously unattainable dual-function performance under the same resource constraints. The impact of IRS-specific limitations, e.g., discrete phase shifts, sparse deployment, on the optimality of trade-offs also deserves closer investigation. Establishing these fundamental relationships is crucial for moving from algorithmic heuristics toward performance-guaranteed IRS-ISAC designs.
\begin{figure*}[!t]
	\centering
	\includegraphics[width=0.8\textwidth]{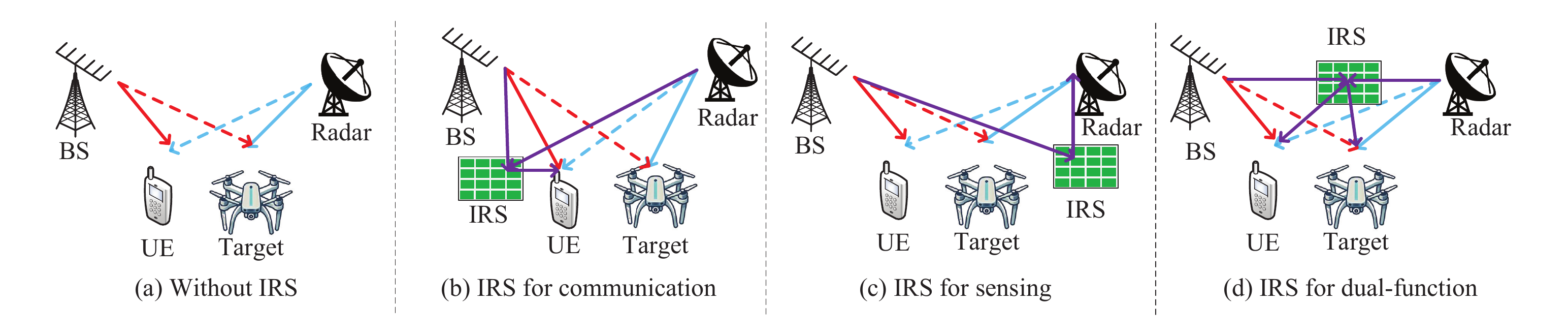}
    \vspace{-10pt}
	\caption{IRS for radar-communication coexistence.}
	\label{rcc}
\end{figure*}
\subsection{Radar-Communication Coexistence} RCC, wherein radar and communication systems independently operate but share the same spectral resources, has emerged as a practical approach to meet the growing spectrum demand in modern wireless networks. Unlike fully integrated dual-function radar-communication (DFRC) systems, RCC maintains hardware independence between radar and communication units but faces the critical challenge of mutual interference, significantly degrading system performance. To effectively manage such interference, recent studies have explored introducing IRS into RCC frameworks. Specifically tailored IRS deployments and advanced beamforming techniques provide unique capabilities to selectively reinforce desired signals and suppress interference, thereby facilitating efficient coexistence between radar sensing and wireless communication operations. This section surveys recent progress in IRS-assisted RCC systems, highlighting key design challenges, innovative algorithmic solutions, and optimal deployment strategies as identified in recent literature.

\subsubsection{Mutual interference mitigation} The concurrent operation of radar and communication systems within shared frequency bands inherently generates mutual interference, a critical challenge that can significantly degrade the performance of both functionalities. As shown in Fig. \ref{rcc}(a), the communication signal interferes with the sensing target, while the sensing signal interferes with the communication user. These interference links are represented by the dashed lines.. By meticulously coordinating these phase shifts across the surface, the IRS can reshape the reflected wavefronts. This capability allows for the constructive interference of desired signals at the intended receiver and, crucially for RCC, the destructive interference of undesired signals, thereby suppressing interference \cite{10130707,10194901,9729741,9264225,  10844061}. 

\begin{figure}
\centerline{\includegraphics[width=0.35\textwidth]{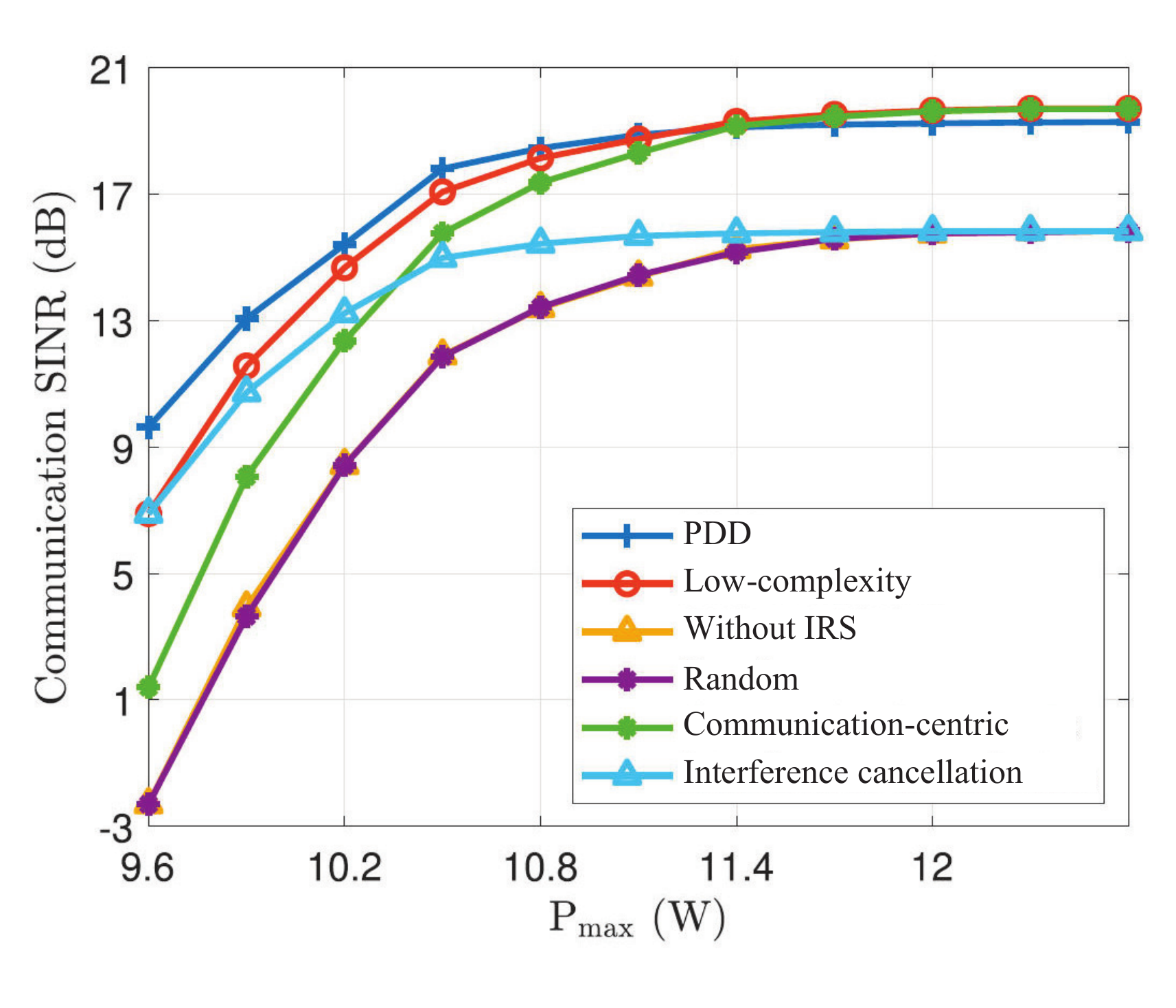}}
\vspace{-5pt}
\caption{Communication SINR versus total radar transmit power in IRS-aided RCC system.}	\label{rcc_sim}
\label{DFRC}
\vspace{-10pt}
\end{figure}

To illustrate the performance gain of IRS in RCC systems, in \cite{9729741}, Fig. \ref{rcc_sim} shows the simulation results for a double-IRS-assisted communication radar coexistence system. In this system, the objective is to jointly optimize the active and passive beamforming to maximize the communication SINR while guaranteeing the radar detection performance (e.g., radar SINR). The simulation results, as shown in Fig. \ref{rcc_sim}, compare the communication SINR versus the total radar transmit power $P_{max}$ for different schemes.
As can be seen from the figure, the IRS-assisted system adopts joint beamforming optimization algorithms (e.g., the PDD (penalty
dual decomposition) and Low-complexity curves) achieves a communication SINR significantly superior to that of the conventional system without IRS (the Without IRS curve). This also verifies the role of the IRS in mitigating interference and enhancing signals. Meanwhile, the performance of the random phase design (the Random curve) is similar to that of the conventional system, which indicates that intelligent beamforming optimization for the IRS is key to achieving performance improvement. In addition, as the radar $P_{max}$ increases, the power available for suppressing interference to the communication receiver increases, and thus the communication SINR of all schemes improves accordingly.

In RCC systems, IRS deployment can be strategically leveraged to mitigate interference along the most detrimental paths, primarily the link from the communication BS to the radar receiver and the link from the radar transmitter to the user equipments (UEs). To mitigate this issue, the IRS can be deployed either at the UE side to suppress sensing interference or at the radar side to cancel communication interference \cite{10263780,9729741}. Alternatively, a balanced strategy involves placing the IRS between the UE and the radar to minimize the mutual interference for both systems \cite{9264225,10844061}. The core principle of IRS-based interference suppression relies on engineering the phase shifts such that the signal reflected by the IRS arrives at the victim receiver (either the radar or the UE) with an opposite phase compared to the signal arriving via the direct interference path. This phase cancellation leads to destructive interference, effectively creating spatial nulls in the interference pattern towards the victim receiver or originating from the interference source. Several studies explicitly formulate optimization problems with the objective of minimizing the interference power generated by the communication system that impinges on the radar receiver \cite{9729741,10844061}. Often, constraints on the maximum permissible interference power at the radar are incorporated into the system design framework to ensure the radar's operational integrity is not compromised \cite{10130707,9264225}. The introduction of IRS provides substantial additional DoFs for spatial filtering compared to conventional active beamforming alone. This enhanced spatial control allows for finer interference management, proving particularly advantageous in scenarios where direct interference paths are weak due to distance or blocked by obstacles \cite{10194901}. Furthermore, advanced architectures employing multiple IRSs, such as deploying one IRS near the communication transmitter and another near the receiver, offer potentially superior interference suppression. This dual-IRS setup enables control over reflections at both ends of the link, allowing for simultaneous mitigation of outgoing interference from the transmitter (protecting the radar) and incoming interference towards the receiver (protecting the UEs) \cite{9729741}. The ultimate effectiveness of IRS in suppressing interference is fundamentally dependent on the accurate acquisition of channel state information (CSI) for all relevant links (direct and IRS-reflected) and the precise, often joint, optimization of the IRS's passive phase shifts alongside the active beamforming strategies employed at the BS and/or the radar transmitter.
\subsubsection{Optimization techniques for joint transmit and reflection beamforming} Integrating IRS into RCC systems introduces a complex optimization challenge: the need to jointly design the active transmission strategies (beamforming, precoding, waveform covariance) at the radar and communication transmitters, concurrently with the passive beamforming (phase shift configuration) at the IRS elements. This joint optimization must navigate the inherent trade-offs between maximizing communication system performance metrics (e.g., sum-rate, weighted sum-rate (WSR), SINR) and ensuring the radar system achieves its required sensing performance (e.g., maximizing DP, maximizing radar SINR, matching a desired beampattern). The problem is further complicated by the tight coupling between the active and passive variables and the non-convex unit-modulus constraint imposed on the reflection coefficients of passive IRS elements, making the overall optimization problem highly non-convex and generally difficult to solve for a global optimum \cite{10130707,10613843,10194901,10263780,9729741,9264225,10844061}.

A dominant strategy to address this complexity involves iterative optimization frameworks, most commonly AO or block coordinate descent (BCD). These methods decompose the intractable joint problem into several more manageable subproblems. In each iteration, one subset of variables (e.g., IRS phase shifts) is optimized while the others are held constant, cycling through the variable blocks until a convergence criterion is met. Within this iterative structure, various optimization techniques are applied to the specific subproblems. For optimizing the active beamformers at the BS or radar, or the transmit covariance matrices, techniques such as weighted minimum mean square error (WMMSE) are used, often to convert sum-rate maximization problems into more tractable forms \cite{10613843}; manifold optimization leverages the geometric structure of the constraints (e.g., constant modulus power) for efficient optimization \cite{10130707}; second-order cone programming (SOCP) is employed when constraints or objectives can be formulated as conic inequalities \cite{10613843}; and Lagrangian dual methods are used to handle constraints and find optimal power or beamforming solutions \cite{10194901}. For optimizing the passive IRS phase shifts, a diverse set of methods has been explored: manifold optimization directly handles the unit-modulus constraint on Riemannian manifolds \cite{10130707}; fractional programming (FP) techniques address sum-of-ratios problems arising in SINR or rate optimization \cite{10613843,10130707}; semidefinite relaxation (SDR) relaxes the non-convex quadratic constraints into convex semidefinite programs, often followed by randomization or rank-one approximation techniques \cite{10613843,10844061}; penalty dual decomposition (PDD) methods handle complex constraints by incorporating them into the objective via penalty terms and dual variables \cite{9729741}; simple local search algorithms provide low-complexity heuristics \cite{10194901}; and more recently, machine learning paradigms, including deep reinforcement learning (DRL) and meta-reinforcement learning (MRL), have been proposed to tackle the high dimensionality and adapt to dynamic channel conditions, potentially reducing computational overhead during inference \cite{10263780}.

The specific optimization objectives pursued reflect diverse system priorities. Goals include maximizing communication sum-rate \cite{10130707}, maximizing WSR while ensuring the radar beam pattern approximates a desired shape \cite{10613843}, maximizing radar SINR or DP subject to communication quality of service (QoS) constraints (e.g., minimum user SINR) \cite{10194901,9264225,10844061}, maximizing communication SINR while guaranteeing radar performance thresholds \cite{9729741}, or maximizing mutual information as a measure of overall system efficiency \cite{10263780}. System architectures investigated range from single IRS deployments, where the IRS is strategically positioned near the transmitter, receiver, or radar to maximize its impact, to more complex double-IRS systems designed for enhanced interference management by controlling reflections at multiple points in the network \cite{9729741}. Additionally, research has considered the optimization of the physical placement of the IRS itself, aiming to maximize coverage within radar exclusion zones or minimize overall system interference \cite{9750859}.

\subsubsection{Practical system considerations}
Transitioning IRS-assisted RCC systems from theoretical models to practical deployments requires addressing several real-world constraints and challenges. While many studies initially assume ideal conditions such as continuous reflection coefficients and perfect CSI, practical hardware and operational limitations necessitate more realistic considerations. A primary factor is the hardware implementation of IRS elements. Due to cost and complexity, most practical IRS implementations utilize components like PIN diodes, which can only provide a finite set of discrete reflection coefficients or phase shifts, rather than continuous tuning \cite{10844061}. Optimizing discrete reflection
coefficients is inherently more difficult than optimizing continuous reflection coefficients because the discrete nature prevents the use of standard gradient-based optimization algorithms. Common heuristic approaches involve optimizing for continuous phases first and then quantizing or rounding the result to the nearest available discrete state \cite{10194901}. However, this simple rounding can lead to significant performance degradation, particularly for interference cancellation tasks where precise phase alignment is critical, as the nearest discrete phase might inadvertently enhance rather than suppress interference \cite{10844061}. More sophisticated techniques have been developed specifically for discrete reflection coefficients optimization, including SDR followed by Gaussian randomization procedures to generate high-quality feasible discrete solutions \cite{10844061}, and localized search algorithms that explore a limited set of discrete neighbors around an initial solution \cite{10613843}. Simulation results indicate that the performance gap between systems using discrete reflection coefficients and those assuming continuous reflection coefficients diminishes as the number of quantization bits (i.e., the resolution of the discrete phases) increases, with 3-4 bits often providing performance close to the continuous case \cite{10613843,10844061}.

Another critical practical challenge lies in acquiring accurate and timely CSI for all relevant propagation paths, including the direct links and the cascaded links involving the IRS (e.g., BS-IRS, IRS-User, IRS-Radar) \cite{10194901}. Since IRS elements are typically passive, they cannot transmit pilot signals themselves, complicating standard channel estimation protocols. This necessitates the development of advanced estimation techniques tailored for IRS-assisted systems, which might involve multi-stage estimation or exploiting specific channel properties. While many foundational studies assume the availability of perfect CSI to establish performance benchmarks \cite{10130707,10613843,10263780,9729741,9264225,10844061}, robust system design must account for inevitable CSI imperfections arising from estimation errors, feedback delays, or channel aging. Research has explored robust optimization frameworks using statistical CSI models, which incorporate knowledge of channel statistics (mean, covariance) rather than instantaneous values, to design beamforming strategies that are less sensitive to channel uncertainty, particularly for the challenging IRS-related links \cite{10194901}.

The physical location of the IRS within the network topology also profoundly influences system performance. An optimally positioned IRS can significantly amplify its benefits in terms of signal enhancement and interference suppression compared to a poorly placed one. Determining the optimal location depends on a multitude of factors, including the spatial distribution of communication users, the relative positions of the BS and radar, the presence and nature of physical obstacles in the environment, and the specific performance objectives of the system (e.g., maximizing coverage probability for users located within a radar exclusion zone) \cite{9750859}. Analytical models, sometimes simplified (e.g., using one-dimensional geometry for tractability), along with optimization algorithms, have been developed to find optimal or near-optimal IRS placements tailored to specific scenarios.

Finally, the computational complexity associated with the joint optimization algorithms remains a significant practical hurdle. Techniques like SDR, PDD, and particularly machine learning approaches like MRL can demand substantial computational resources, especially as the number of antennas at the BS/radar and the number of IRS elements scale up \cite{10613843,10194901,10263780,9729741,10844061
}. Consequently, a key research direction involves the development of low-complexity algorithms. This can be achieved by exploiting specific structural properties of the problem, designing algorithms tailored to particular operating regimes (e.g., distinguishing between high-power and low-power radar scenarios \cite{9729741}, or assuming communication-centric operation \cite{10844061}), or leveraging efficient learning frameworks like MRL that aim to reduce training overhead and enable faster adaptation \cite{10263780}. Balancing performance optimality with computational feasibility is crucial for the practical deployment of IRS in future RCC systems.

\begin{figure}
\centerline{\includegraphics[width=0.65\linewidth]{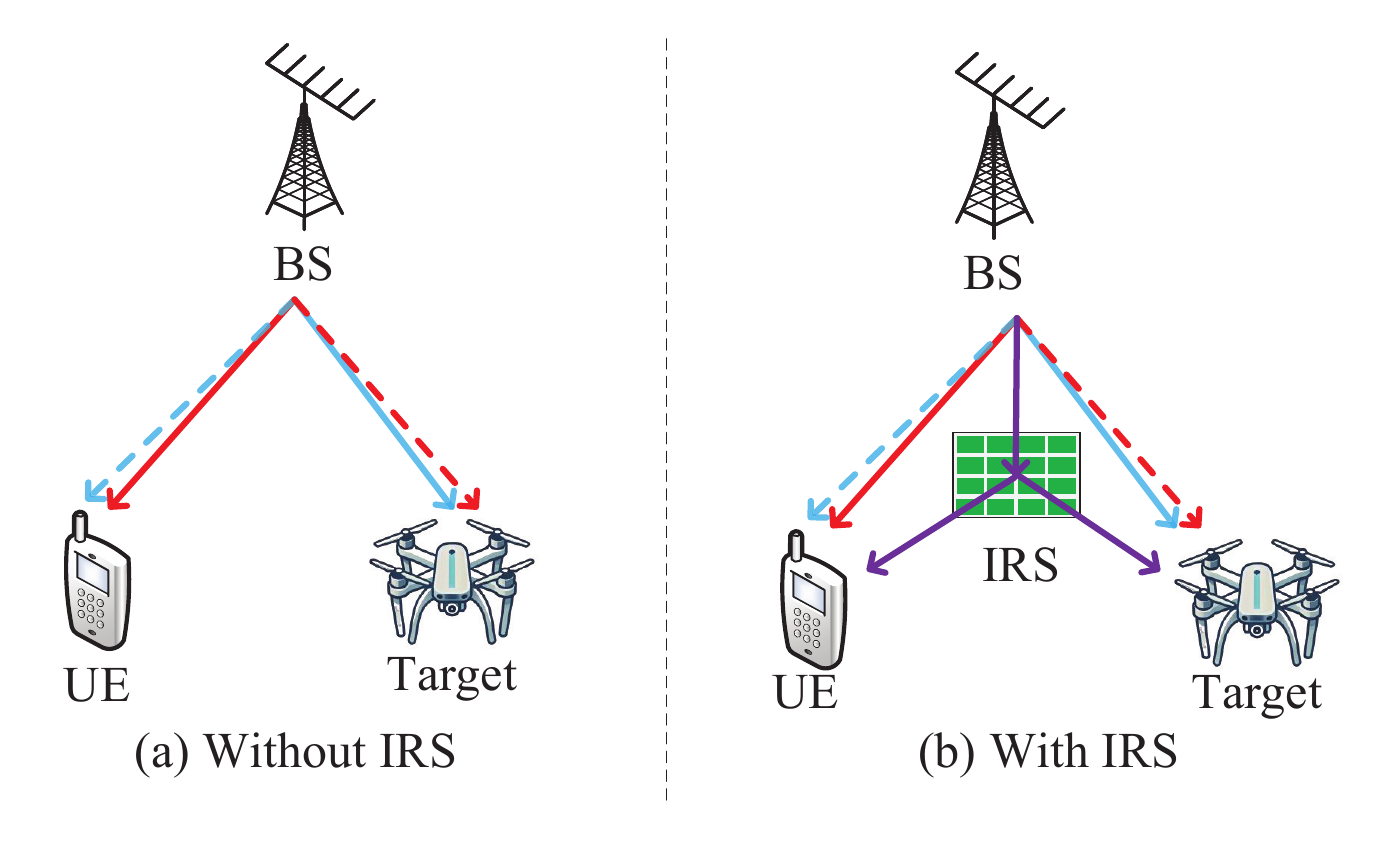}}
\vspace{-10pt}
\caption{IRS for dual-functional radar-communication.}
\label{DFRC}
\vspace{-10pt}
\end{figure}
\subsection{Dual-Functional Radar-Communications} The convergence of sensing and communication functionalities within the same wireless infrastructure is a defining feature of 6G networks. DFRC systems are envisioned to enable this convergence by utilizing a common set of waveform and hardware resources. However, realizing that DFRC faces inherent trade-offs between sensing accuracy and communication reliability. Fig. \ref{DFRC}(a) illustrates that a dual-functional BS partitions its transmit resources for communication (blue lines) and sensing (red lines). Nevertheless, the broadcast characteristic of the signals inherently causes mutual interference between the two systems. IRS, with its ability to passively manipulate signal propagation, offers new DoFs to control these trade-offs. As shown in Fig. \ref{DFRC}(b), in the context of DFRC, IRS can steer reflected beams to reinforce desired communication directions while simultaneously shaping radar beampatterns to enhance target detection or suppress interference. In what follows, we highlight several key research directions that have emerged around IRS-assisted DFRC design.

\begin{table*}[htbp]
\centering
\renewcommand{\arraystretch}{1.35}
\footnotesize
\caption{IRS-DFRC}
\begin{tabular}{| p{2.2cm} | p{4cm} | p{2.5cm} | p{3.5cm} | p{3.2cm}  |}
\hline
\textbf{Technical Theme} & \textbf{Key Focus} & \textbf{Optimization Methods} & \textbf{Strengths} & \textbf{Challenges} \\
\hline
Beamforming for Dual-Functionality \newline \cite{10052711, 9769997, 9416177, 9852716} &
Joint BS active and IRS passive beamforming/waveform design for ISAC performance trade-offs &
AO/BCD, MM, SDR, MO, FP &
ISAC synergy; Enhanced signal/coverage; Resource efficiency. &
Non-convexity; Coupled variables; Performance trade-offs. \\
\hline
Sensing-Centric Design \newline \cite{10141975, 10440056, 10200033, 10384870, liu2023snr, 10938377} &
Prioritizing sensing metrics (SNR, CRB, Pd, Beampattern) under communication QoS constraints. &
AO/BCD, SDR, MM, MO &
Improved detection/estimation accuracy; Clutter suppression. &
Limited comm. performance; Sensitivity to models/targets. \\
\hline
Sensing-Assisted Communication \newline \cite{10632049, 10436416, 9966507, 10197455, 10521619} &
Using sensing outputs (location, velocity) to improve communication beamforming/resource allocation/modulation. &
Location-aware Optimization, AO &
Accurate beamforming with imperfect CSI; Context-aware resource use &
Requires accurate/fast sensing; Error propagation. \\
\hline
Multi-IRS Collaboration \newline \cite{10695883, 10475369, 10411853, Net26_CHuang_TWC2025} &
Using multiple distributed IRSs for extended coverage, more DoFs, and better localization. &
Distributed AO/BCD, SDR, Stochastic Geometry &
Extended coverage/reliability; Enhanced optimization; Improved localization accuracy. &
System complexity; Coordination overhead; Synchronization. \\
\hline
System Constraints and Robust Design \newline \cite{10443608, 10141975, 10143420, 10056405, 9591331, 10922191} &
Designing for robustness against imperfect CSI, hardware limits (discrete phases), clutter, security threats. &
Robust Optimization, Secure BF, SDR &
Reliability in practice; Handles uncertainties; Enhanced security. &
Performance vs. robustness trade-off; Complex modeling. \\
\hline
\end{tabular}
\end{table*}

\subsubsection{IRS-aided beamforming for dual-functionality} A primary obstacle hindering the performance of ISAC systems is signal blockage, particularly prevalent in complex propagation environments and at higher frequency bands like mmWave. Passive IRS provides a powerful mechanism to counteract this limitation by intelligently redirecting signals around obstructions \cite{10141975, 10440056, 10143420, 9364358}. By meticulously controlling the phase shifts of its constituent elements, an IRS can establish reliable virtual LoS propagation paths. This capability is crucial for extending service coverage to communication users situated in shadowed areas and, critically, for enabling the sensing of targets that lack a direct LoS path to the BS \cite{10695883, Net26_CHuang_TWC2025}. The creation of these controllable reflective links essentially reconfigures the wireless channel itself to overcome the physical barrier posed by blockages.

Beyond overcoming blockages, passive IRS actively combats the challenge of signal attenuation due to path loss. While IRS enhances received signal strength for communication users through coherent superposition of reflected signals \cite{10632049, 10042240}, its role in sensing is particularly significant in addressing the severe path loss associated with the double reflection (BS-IRS-Target-IRS-BS) typically required for detecting NLoS targets \cite{10042240, 10440056}. By optimizing its phase shifts, the IRS can focus the reflected energy towards the target during illumination and coherently combine the weak echo signals reflected back towards the BS, thereby boosting the radar SNR and improving target detection and estimation performance despite the challenging propagation path \cite{10042240, 10200033, 9364358}. Some studies also explore how the IRS-induced multipath environment can be leveraged to provide spatial diversity gain, mitigating the detrimental effects of target radar RCS fluctuations \cite{10564104}.

\begin{figure}
\centerline{\includegraphics[width=0.35\textwidth]{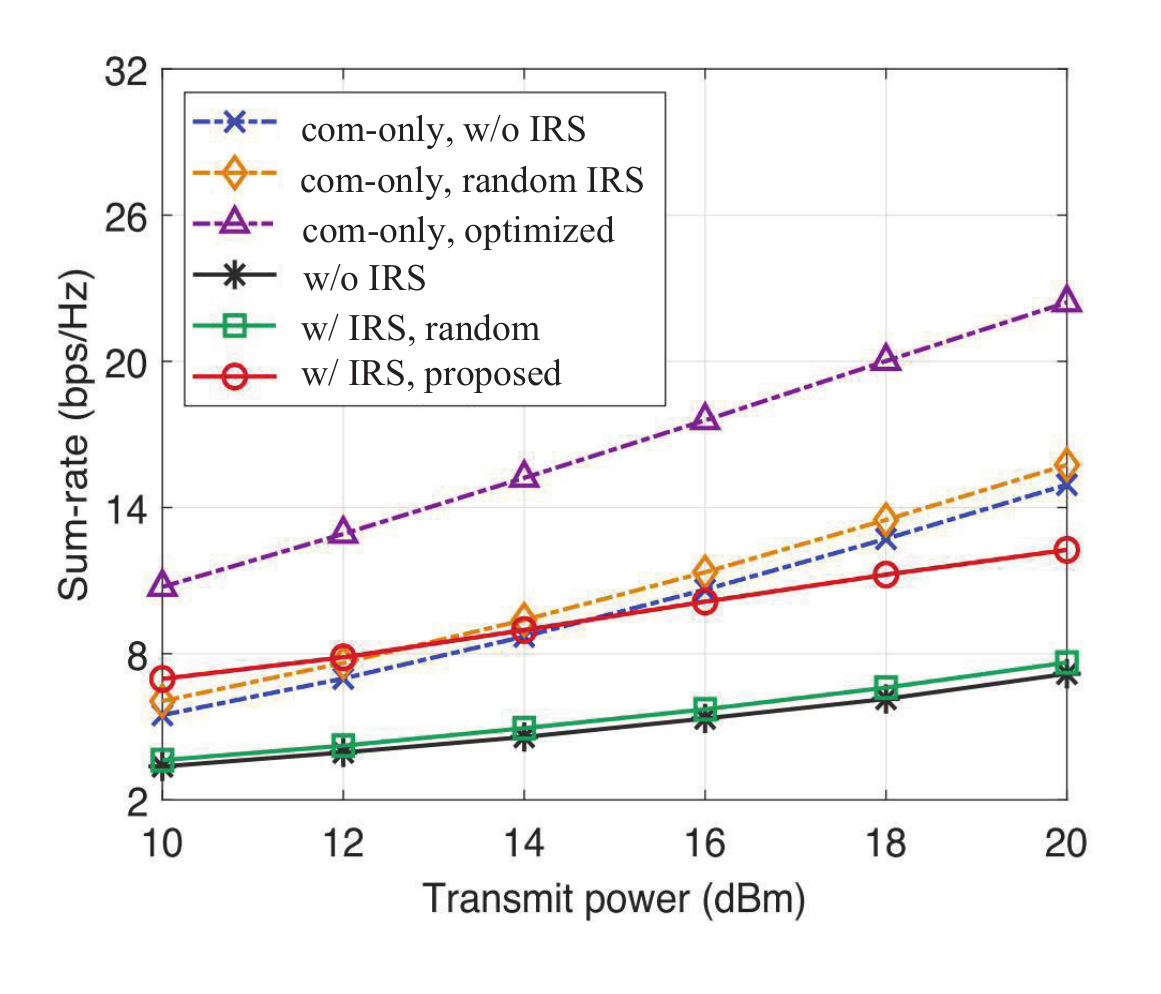}}
\vspace{-5pt}
	\caption{Sum-rate versus transmit power in DFRC system.}
	\label{DFRC_fig}
    \vspace{-10pt}
\end{figure}

The inherent integration of dual functions in ISAC systems presents the difficulty of managing performance trade-offs between sensing and communication objectives. Passive IRS introduces a significant number of additional DoFs through its adjustable phase shifts, offering enhanced flexibility in navigating these trade-offs \cite{10042240, 9769997, 9852716}. Through joint beamforming design, encompassing the BS's active beamforming and the IRS's passive beamforming, these DoFs can be exploited to optimize system performance strategically. For example, systems can be designed to maximize communication metrics like sum-rate while strictly satisfying sensing requirements defined by radar beampattern similarity \cite{9416177, 9852716} or CRB constraints \cite{10440056}. 
To demonstrate the communication enhancement by IRS in ISAC, in \cite{9852716}, Fig. \ref{DFRC_fig} simulates an IRS-assisted ISAC system. It maximizes the communication sum-rate under a radar sensing beampattern similarity constraint via joint optimization of BS active and IRS passive beamforming. As shown in Fig. \ref{DFRC_fig}, deploying an IRS (``w/ IRS, ...'' schemes) improves communication performance compared to the no IRS case (``w/o IRS'') by introducing additional NLoS links to enhance downlink communication. More significantly, the jointly optimized IRS scheme (``w/ IRS, proposed'') substantially outperforms the IRS scheme with random phases (``w/ IRS, random''), clearly indicating the necessity and effectiveness of the joint optimization design. Furthermore, comparing the ISAC system (e.g., ``w/ IRS, proposed'') with the corresponding communication-only system (``com-only, optimized'') reveals a performance gap, illustrating the communication-sensing trade-off due to satisfying the sensing constraint.

Conversely, sensing performance, such as radar SNR \cite{10042240, 10200033} or DP \cite{10384870, 10042425}, can be maximized under constraints guaranteeing minimum communication QoS for multiple users \cite{9769997, 10254508}. The IRS essentially provides an additional layer of control to fine-tune the balance between the two functions.

Furthermore, passive IRS can help alleviate performance degradation caused by practical hardware constraints at the BS, such as the use of power-efficient constant modulus waveforms. While constant modulus waveforms limit the flexibility of active beamforming, the spatial control offered by the IRS can compensate, allowing the system to meet ISAC objectives even with restricted waveform choices \cite{10042240, 9769997, 9416177, 9591331}. Similarly, joint optimization frameworks strive to harness the IRS's benefits even when accounting for practical discrete phase shifts at the IRS elements \cite{10443608, 9591331}. The effectiveness of IRS in resolving these diverse DFRC challenges fundamentally relies on the co-optimization of BS transmission parameters and IRS reflection coefficients, demanding sophisticated design methodologies explored in subsequent sections.

\subsubsection{Sensing-centric design} While ISAC aims for synergy, scenarios often arise where prioritizing the sensing function is paramount, necessitating designs specifically tailored to maximize sensing performance under communication constraints. Passive IRS plays a crucial role in these sensing-centric designs by offering enhanced control over signal propagation to overcome inherent sensing challenges, such as detecting weak targets or achieving high estimation accuracy, especially in NLoS conditions \cite{10440056,10200033}. The core approach involves jointly optimizing the BS active beamforming and waveform with the IRS passive phase shifts to explicitly favor sensing objectives. One primary difficulty in ISAC sensing is ensuring sufficient energy illuminates the target and that the weak reflected echo can be reliably detected at the BS, particularly when using passive IRS, which introduces double reflection path loss \cite{10052711, 10440056}. Sensing-centric designs address this by maximizing metrics directly related to detection performance. This often involves optimizing the joint BS-IRS beamforming to maximize the radar SNR at the receiver \cite{10052711, 10200033, 9769997} or maximizing the transmit beampattern gain towards the assumed target location \cite{10141975, 10143420, 10122520}. By carefully shaping the wavefronts via the IRS, signal energy can be concentrated in the target's direction, significantly improving the probability of detection, even for NLoS targets or targets embedded in clutter \cite{10141975, 10200033, 10384870}. Maximizing the DP itself, sometimes based on the target's estimated scattering characteristics, can also be the direct objective \cite{10384870}. Beyond mere detection, achieving high accuracy in estimating target parameters (like location, velocity, or DoA) is often critical. Passive IRS contributes significantly here by providing controllable reflected paths that enhance spatial resolution. Sensing-centric designs focused on estimation accuracy typically aim to minimize the CRB, which represents a lower bound on the variance of any unbiased estimator \cite{10440056, 10122520, 10527368, 9591331, 10938377}. Jointly optimizing the BS transmit strategy and IRS phase shifts to minimize the CRB for target DoA or position allows the system to achieve finer sensing precision than possible without the IRS, especially under constraints like discrete phase shifts \cite{9591331} or when dealing with multiple targets \cite{10938377}. Another approach quantifies sensing performance using information-theoretic metrics. Maximizing the mutual information between the received echo signal and the target's response or parameters provides a measure of the information gain achieved through sensing \cite{9844707, li2024joint, 10056405}. Sensing-centric designs based on mutual information (MI) maximization aim to configure the BS transmission and IRS reflection to capture the maximum possible information about the target scene, subject to ensuring necessary communication quality for simultaneous users \cite{9844707, 10056405}. These designs often reveal fundamental trade-offs between allocating resources for maximizing sensing information versus communication throughput. Crucially, all these sensing-centric optimization frameworks must operate within the constraints imposed by the concurrent communication task. The maximization of radar SNR, minimization of CRB, or maximization of sensing MI is typically subject to constraints guaranteeing minimum data rates or SINRs for communication users \cite{10440056, 10200033, 10384870, 9769997, 10122520, 10938377}. The passive IRS, through the additional DoFs it offers in the joint beamforming design, provides vital flexibility to meet these communication requirements while pushing the boundaries of sensing performance.

\subsubsection{Communication-centric design} 
The communication-centric design paradigm explicitly prioritizes communication performance, with its core idea being the integration of sensing functionalities onto existing communication platforms. In specific system optimization problems, design philosophy is manifested by treating communication performance metrics as the primary optimization objective, while sensing performance is handled as a constraint that must be satisfied. For instance, some research focuses on the joint optimization of the transmit beamforming at the BS and the phase shifts of the IRS to maximize the system's communication sum-rate, while ensuring that sensing efficiency does not fall below a predetermined threshold \cite{10160136}. Similarly, as energy efficiency is a key performance indicator for communication systems, other studies aim to maximize it through joint BS and IRS beamforming design, subject to dual constraints on both communication rate and sensing quality. Furthermore, some works employ a weighted-sum approach to balance the dual functionalities, reflecting the priority of communication through the adjustment of weighting factors \cite{9844707}. For example, one study achieves joint optimization by maximizing a weighted sum of the communication data rate and the sensing mutual information, while another formulates a weighted objective function by jointly minimizing MU interference for the communication side and maximizing the radar SINR \cite{10042240}.

Another design paradigm is to use perceived information to improve the performance of communication systems. This paradigm directly addresses the difficulty of acquiring accurate and timely CSI or user location data for communication optimization, particularly in dynamic environments or non-NLoS scenarios facilitated by passive Reconfigurable Intelligent Surfaces IRS \cite{10436416, 10197455}. The IRS plays a dual role here: first, aiding the initial sensing phase to gather more reliable environmental information, and second, being configured based on this sensed information to improve the subsequent communication link.

A primary way sensing assists communication is through location-aware beamforming. Accurate knowledge of user positions, obtained via IRS-aided sensing mechanisms, allows the BS and the IRS to collaboratively steer communication beams more precisely towards intended users \cite{9966507,10197455}. This is particularly beneficial when traditional pilot-based channel estimation is difficult or incurs high overhead, such as for the cascaded BS-IRS-user channel. By using sensed location information (e.g., angles) as a basis, joint active and passive beamformers can be designed to maximize the signal energy delivered to the estimated user location, thereby improving communication SNR and data rates even with imperfect or outdated CSI \cite{9966507,10197455}. Some protocols explicitly employ a two-phase approach, where coarse sensing in the first phase informs the beamforming design for enhanced communication and potentially finer sensing in the second phase \cite{10197455}.

Sensing outputs can also inform more efficient radio resource allocation for communication. Information gathered about the environment, such as user locations relative to the BS and IRS, or potentially even inferred channel conditions, can guide the allocation of time slots, power, or frequency resources among users \cite{10436416}. For instance, joint optimization of time allocation between dedicated sensing periods and communication periods, along with IRS reflecting precoding design, can maximize the overall communication capacity based on the quality of sensing results obtained, while managing interference effectively \cite{10436416}. This context awareness derived from sensing enables a more adaptive and efficient utilization of shared ISAC resources.

Furthermore, sensing information can enable novel communication strategies mediated by the IRS. The sensed location or angle of a user relative to the IRS can be directly incorporated into the design of the IRS reflection pattern itself. One such approach involves IRS-based space-time coding, where the time-varying reflection pattern of the IRS is designed based on the sensed user angle to simultaneously perform modulation and beamforming, effectively embedding communication symbols onto the reflected sensing waveform \cite{10521619}. This integrates sensing results directly into the physical layer communication mechanism via the IRS configuration. Similarly, in high-mobility scenarios employing Orthogonal Time Frequency Space (OTFS) modulation, user velocity estimated through sensing can be used to optimize the joint BS-IRS beamforming in the delay-Doppler domain, mitigating mobility-induced impairments and enhancing communication reliability \cite{10632049}. This demonstrates a tighter integration where sensing directly refines the parameters of advanced communication schemes.

\begin{figure*}[!t]
	\centering
	\includegraphics[width=0.7\textwidth]{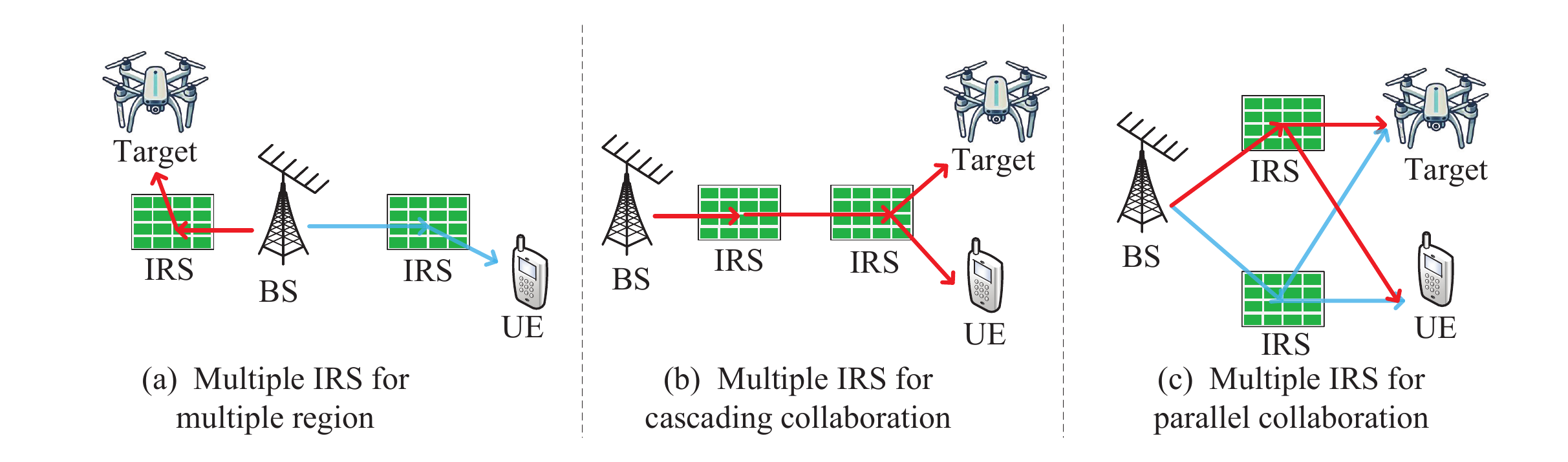}
	\caption{Multiple IRSs for ISAC.}
	\label{multi}
    \vspace{-5pt}
\end{figure*}

\subsubsection{Multi-IRS cooperative mechanisms} While a single IRS offers significant local control over the radio environment, its impact might be insufficient for ISAC systems operating over larger geographical areas or facing complex blockage scenarios. The deployment of multiple passive IRSs, acting in collaboration, presents a viable solution to overcome the inherent limitations of a single surface, primarily by extending coverage and providing enhanced spatial control \cite{10695883, 10411853, 10218356}. This collaborative approach aims to create a more pervasive and finely-tunable smart radio environment.

A key difficulty addressed by multi-IRS systems is the limited coverage range of a single IRS. As shown in Fig. \ref{multi}(a), by strategically distributing multiple passive IRSs, perhaps across different building facades or lamp posts, it becomes possible to serve UEs and sense targets spread across significantly larger areas or located within distinct shadow zones where direct BS signals cannot reach \cite{10695883, Net26_CHuang_TWC2025}. This distributed architecture allows for the establishment of cooperative virtual links, potentially involving multiple IRSs reflecting signals sequentially or in parallel, thereby enhancing signal availability and reliability for both communication and sensing tasks compared to single-IRS deployments \cite{10411853}. As depicted in Fig. \ref{multi}(b), an alternative approach for coverage extension involves arranging multiple IRSs in a series-cascade. This deployment promotes multi-hop signal propagation, allowing the signal to strategically circumvent obstructions. While such a link is subject to increased path loss, it enables the accumulation of a more substantial beamforming gain from the multiple reflections \cite{10643789,10159017}. 

Beyond coverage extension, the collective use of multiple IRSs provides a substantial increase in the total number of controllable reflecting elements. This translates into a much larger pool of spatial DoFs available for joint beamforming optimization across the BS and all participating IRSs \cite{10411853, 10218356} as shown in Fig. \ref{multi}(c). These enhanced DoFs are particularly valuable for tackling complex ISAC scenarios involving multiple users and multiple targets simultaneously. They allow for more precise shaping of radio waves, potentially enabling simultaneous beam focusing towards multiple targets for improved sensing while carefully managing interference towards multiple communication users to meet their QoS demands \cite{10411853}. The ability to coordinate phase shifts across multiple surfaces offers greater flexibility in resolving the sensing-communication performance trade-offs inherent in multifaceted ISAC applications, such as those found in multicell cooperative systems where network-wide power efficiency is critical \cite{10411853}. The optimization challenge in wideband multi-IRS systems also necessitates sophisticated joint designs to handle frequency-selective channels and Doppler effects for moving targets \cite{10218356}.

Multi-IRS collaboration also significantly enhances the system's sensing capabilities, especially for high-accuracy localization. Utilizing two or more spatially separated passive IRSs creates multiple, diverse perspectives for observing a target. This geometric advantage facilitates triangulation methods based on AoA information extracted from the distinct reflection paths associated with each IRS, leading to substantial improvements in localization accuracy, particularly for NLoS targets where direct paths are unavailable \cite{10475369}. Achieving centimeter-level accuracy has been demonstrated in dual-passive-IRS frameworks \cite{10475369}. However, effectively harnessing the benefits of multi-IRS collaboration demands sophisticated coordination strategies. The joint optimization of phase shifts across all IRSs becomes significantly more complex, requiring scalable algorithms to manage the high dimensionality and ensure coherent cooperation while mitigating potential inter-IRS interference \cite{10411853, 10218356}.

\section{IRS-aided ISAC: Mutualism}
Achieving mutual enhancement (mutualism) between communication and sensing (C$\&$S) functions is a significant objective to achieve for ISAC systems. Utilizing communication networks (e.g., signals, infrastructures, protocols) to strengthen sensing capabilities, while in turn using sensing information (e.g., CSI, angle, distance) to promote communication performance, is pivotal to realizing this reciprocity. The introduction of the IRS provides powerful support for implementing these mechanisms. This section focuses on effective C$\&$S enhancement mechanisms empowered by IRS, reviewing research on Sensing-aided Communication (SAC), Communication-aided Sensing (CAS), and C$\&$S Mutualism (CSM), respectively.

\subsection{Sensing-aided Communication (SAC)}
One prominent paradigm to implement SAC is to exploit the latest sensing information (e.g., CSI, angle, distance) and its prediction in beamforming and scheduling design to promote the subsequent communication performance. The authors of \cite{CAS01_ZFei_TWC2024} considered a sensing-assisted predictive beamforming in a vehicle-to-infrastructure (V2I) system. Considering fast varying channels due to high mobility of vehicles, \cite{CAS01_ZFei_TWC2024} proposed a deep learning-based predictive beamforming paradigm to promote communication rate. In the first stage, AoA is updated based on the echo signals and predicted by a dedicated neural network (NN). Then, the transmit beamforming vector at the BS and the reflection pattern of the IRS are jointly optimized for communication. The authors of \cite{SAC02_KTMeng_TVT2024} proposed an intelligent omni-surface (IOS) aided predictive beamforming protocol for a vehicle communication system. The proposed protocol divides each transmission frame into two stages. In the first stage, the IOS mounted on vehicles reflects part of the incident communication signals back to the roadside unit (RSU), which updates angle and distance information via sensing, predicts the channel of the forthcoming frame through extended Kalman filtering (EKF), and adjusts beamforming in awareness of CSI prediction errors. In the second stage, IOS refracts the entire incident communication signals to users to promote the communication rate. In \cite{SAC18_XueyanCao_TVT2025}, the authors considered predictive beamforming in a V2X ISAC system aided by distributed IRSs. According to \cite{SAC18_XueyanCao_TVT2025}, in each coherence period, an initial pilot-based location phase is conducted to acquire the accurate location of the mobile user. After that, the BS continually transmits communication data, tracking and predicting the user's location status through EKF, and periodically adjusts the transmitting beam. The authors of \cite{SAC03_RWang_TComm2023} proposed an IRS-aided simultaneous localization and communication (SLAC) scheme in an mmWave transmission system. Specifically, in the downlink transmission, the BS first transmits pilots to acquire the position and orientation information of the mobile user. Then, based on the newly obtained location information, the BS jointly optimizes sensing time allocation and beamforming before data transmission to improve the achievable communication rate. 

Besides the predictive beamforming schemes previously mentioned, IRS also facilitates diverse SAC functions. The authors of \cite{SAC04_XYWang_WCL2021} considered a sensing-aided transmission scheme in a communication and radar coexistence system aided by an IRS. Specifically, during the sensing stage, the BS and user sequentially transmit pilot signals to acquire CSI, followed by passive and active beamforming to improve both communication and sensing performance. The work \cite{10436416} proposed an IRS-aided spectrum sensing (SS) type transmission scheme. Concretely, during the first stage, the BS first transmits a signal to the intended receiver through both the BS-receiver and BS-IRS-receiver links and detects potential users lying in the direct BS-receiver channel. Depending on the detection result, if there are no users lying in the way to the target receiver, the BS transmits through a direct link. Otherwise, BS communicates through the virtual link by IRS to avoid interference. The authors of \cite{SAC13_JieHu_arXiv2025, SAC14_JieHu_arXiv2025, SAC15_JieHu_CL2025} recently proposed a novel reconfigurable intelligent sensing surface (RISS) architecture, which can sense the arriving angle of incident signals and recognize user identity (ID) of the signals. Leveraging this feature, \cite{SAC13_JieHu_arXiv2025} proposed a sensing-based beamforming strategy. According to the ID recognition result, the RISS performs either constructive beamforming utilizing the sensing AoA to enhance the SNR of the intended user or destructive beamforming to mitigate interference. In \cite{SAC14_JieHu_arXiv2025}, the authors considered a sensing-assisted wirelessly powered communication network (WPCN) that utilizes RISS's sensing capability to facilitate uplink wireless information transfer (WIT) and elevate downlink wireless energy transmission (WET). In \cite{SAC15_JieHu_CL2025}, multiple RISSs are deployed and their placements are optimized to enlarge the sensing area of RISSs.

\begin{table*}[htbp]
\centering
\renewcommand{\arraystretch}{1.25}
\footnotesize
\caption{SAC Summary}
\begin{tabular}{|p{4.2cm}|p{2cm}|p{2.3cm}|p{3.5cm}|p{3.5cm}|}
\hline
\textbf{SAC Scheme} & \textbf{System Setup} & \textbf{Sensing Paradigm} & \textbf{Problems} & \textbf{Solutions} \\
\hline

Sensing stages update and predict CSI (e.g., AoA, AoD, range) to facilitate predictive beamforming 
& V2I, DL, H-IRS, single user
& Echo of DL communication signal
& Estimate and predict AoA; apply predictive beamforming
& AoA prediction NN + beamforming optimization; end-to-end beamforming NN \cite{CAS01_ZFei_TWC2024} \\
\hline

Sensing stages update and predict CSI (e.g., AoA, AoD, range) to facilitate predictive beamforming 
& V2X, DL, IOS, multi-users
& Echo of DL communication signal
& Estimate azimuth and elevation angles; design beamforming
& Asymptotic analysis; AO \cite{SAC02_KTMeng_TVT2024} \\
\hline

Sensing stages update and predict CSI (e.g., AoA, AoD, range) to facilitate predictive beamforming 
& V2X, UL+DL, multi-H-IRSs, single user
& UL pilot signal
& Estimate/predict azimuth and elevation; design beamforming
& 2D-MUSIC + EKF; AO + CCCP \cite{SAC18_XueyanCao_TVT2025} \\
\hline

Sensing stages update and predict CSI (e.g., AoA, AoD, range) to facilitate predictive beamforming 
& DL, IRS, single user
& UL pilot signal
& Joint optimization of time allocation and beamforming to improve CRB and communication rate
& Robust optimization; AO \cite{SAC03_RWang_TComm2023} \\
\hline

Acquiring sensing and communication CSI
& Radar, BS, IRS, one target, multi-users
& UL and DL pilot signals
& Design beamforming to maximize DP while ensuring comm. rate
& AO + SDR \cite{SAC04_XYWang_WCL2021} \\
\hline

Detecting potential users in the way
& BS, IRS, one user, potential interfered user
& Echo of DL communication signal
& Optimize beamforming to enhance detection precision
& AO + convex optimization \cite{10436416} \\
\hline

Acquiring AoA and ID of impinging signals to enable interference-aware beamforming
& BS, DL, RISS, one user, one interfering user
& UL communication signals
& Constructive beamforming for intended user while suppressing interference
& Convex optimization; hardware validation \cite{SAC13_JieHu_arXiv2025} \\
\hline

Acquiring AoA and ID of impinging signals to enable interference-aware beamforming
& BS, RISS, UL WIT users, DL WET users
& UL and DL signals
& Joint beamforming to maximize UL rate and DL WET efficiency
& AO + robust optimization \cite{SAC14_JieHu_arXiv2025} \\
\hline

Acquiring AoA and ID of impinging signals to enable interference-aware beamforming
& BS, DL, one user, multi-RISS
& Echo of DL communication signal
& Beamforming design to enlarge sensing area and improve DL comm. rate
& Convex optimization \cite{SAC15_JieHu_CL2025} \\
\hline

\end{tabular}
\end{table*}

\subsection{Communication-aided Sensing}
To accomplish the conversion of cellular networks from their current communication-centric form into real ISAC systems, various aspects of communication networks, including signals, infrastructures, protocols, and resources, need to be effectively exploited to strengthen sensing capability, especially combined with the deployment of IRS. 


One straightforward method is exploiting IRS to reinforce the classic pilot-based sensing capability with respect to targets and environment, which naturally inherits the channel estimation and positioning functions of 5G NR network. The paper \cite{10552748} investigated user localization and environment sensing through standard uplink channel estimation for IRS-assisted MU mmWave MIMO-OFDM systems. The authors of \cite{10552748} proposed to first utilize a complex-valued depth residual convolution neural network (C-DRCNN) to recover channel coefficients and then to utilize parallel factor (PARAFAC) tensor decomposition to extract CSI parameters of users/scatters, including angle of arrival (AOA), AoD, and time delay (equivalently distance). In \cite{CAS06_YXLin2022}, the authors proposed a novel twin-IRS structure constituted by two planar IRS devices with a predefined relative spatial rotation. By designing IRS-enabled training coefficients and accommodating training signals into a third-order canonical polyadic tensor, \cite{CAS06_YXLin2022} designed an estimator to extract the location parameters of users/scatters embedded in cascaded channels by leveraging sparse signal processing techniques. Besides, joint user localization and environment sensing were shown to be effective for systems employing 3D conformal curved IRSs and circular IRSs, in \cite{CAS04_YXLin2022} and \cite{CAS07_YXLin2024}, respectively, through tensor-based signal processing techniques. 

Besides the aforementioned pilot-based sensing, IRS can also be well incorporated into diverse functions, including measurements, resource allocation, and beam sweeping, to implement perception capability. The authors of \cite{CAS09_BoWang_SenJ2022} considered IRS-assisted NF localization based on the RSS of the target node. By transmitting pilots from the anchor node, adjusting IRS phase shifts, and processing RSS measurements, the work \cite{CAS09_BoWang_SenJ2022} developed a localization method for both LoS and NLoS environments. The authors of \cite{CAS12_HZhang_CL2021} proposed an IRS empowered RSS-based positioning scheme. By designing an IRS reflection pattern, the propagation channels by way of the IRS are designed in a way that the differences between the RSS values of adjacent locations are effectively enlarged, such that the mobile device's position can be accurately identified. The \cite{CAS10_BoWang_TITS2024} considered the SLAC paradigm in an IRS-aided MU multiple-input multiple-output (MISO) Orthogonal Frequency Division Multiplexing (OFDMA) system, where pilot and data signals are allocated on different sub-carriers to implement both C\&S functions. The authors of \cite{CAS10_BoWang_TITS2024} proposed jointly optimizing sub-carrier assignment and beamforming to provide balanced C\&S performance. The work \cite{SAC07_XYCao_TVT2023} proposed an IRS-enabled feedback-based beam training scheme. Considering the coexistence of the communication user and sensing target, the authors of \cite{SAC07_XYCao_TVT2023} proposed a dynamic particle swarm optimization (PSO) based beam sweeping strategy, which utilizes received echo and the communication user's feedback, to effectively promote sensing SNR.

Additionally, mobile users' cooperation is indeed a unique advantage of cellular networks to implement diverse sensing functions, especially assisted by IRS. In \cite{CAS08_XYuan_TWC2023}, the authors considered joint user localization and information recovery in an IRS-enabled reflection modulation system. Specifically, IRS is mounted on each mobile device to reflect the incident signal back to the BS with modulated information. Based on the received echo from mobile users, the BS conducts a message-passing algorithm on a factor graph to retrieve both uplink data and the location of each mobile user. In \cite{CAS05_YXLin2025}, the authors proposed a novel self-localization paradigm in an MU network empowered by a 3D spherical IRS. In this paradigm, active users transmit pilot or data signals that are redirected by a spherical IRS with designed profiles to all users. The authors developed an estimator based on subspace signal processing techniques to extract the channel parameters and accomplish self-localization of mobile users. The authors of \cite{CAS11_MMZhao_ISWCS2022} studied IRS-aided localization in an NF MU communication system, where the location of users is acquired through signals not only from the IRS but also from cooperating mobile users. It was verified in \cite{CAS11_MMZhao_ISWCS2022} that, via optimizing power allocation and IRS phase shifts, the location accuracy can be effectively improved.

\begin{table*}[htbp]
\centering
\renewcommand{\arraystretch}{1.25}
\footnotesize
\caption{CAS Summary}
\begin{tabular}{|p{2.8cm}|p{3.8cm}|p{2.6cm}|p{3.2cm}|p{3.5cm}|}
\hline
\textbf{CAS Scheme} & \textbf{System Setup} & \textbf{Sensing Paradigm} & \textbf{Problems} & \textbf{Solutions} \\
\hline

\multirow{4}{=}{\parbox[t]{2.8cm}{\raggedright Pilot-based channel estimation and localization}}
& MIMO-OFDM, UL, IRS, multi-users & UL pilot
& Estimate azimuth and elevation angles of BS-IRS and IRS-user channels
& CNN + tensor decomposition \cite{10552748} \\
\cline{2-5}
& MIMO-OFDM, UL, twin-IRS, multi-users & UL pilot
& Estimate azimuth and elevation angles of BS-IRS and IRS-user channels
& Atomic norm optimization, tensor decomposition \cite{CAS06_YXLin2022} \\
\cline{2-5}
& MIMO-OFDM, conformal IRSs, multi-users & UL pilot
& Estimate angles and location coordinates of users/scatters
& ESPRIT, solving nonlinear equations \cite{CAS04_YXLin2022} \\
\cline{2-5}
& MISO-OFDM, circular IRS, multi-users & UL pilot
& Estimate angles and location coordinates of users/scatters
& Subspace estimation, tensor theory \cite{CAS07_YXLin2024} \\
\hline

\multirow{2}{=}{\parbox[t]{2.8cm}{\raggedright RSS measurement-based localization}}
& 2D, NF, anchor node, IRS, target & Pilot from anchor
& Estimate target's location
& Weighted least squares \cite{CAS09_BoWang_SenJ2022} \\
\cline{2-5}
& AP, multi-IRSs, DL, multi-users & UL pilot
& Design IRS phase shifts to enlarge RSS values of adjacent locations
& Gradient descent + local search \cite{CAS12_HZhang_CL2021} \\
\hline

\parbox[t]{2.8cm}{\raggedright OFDMA-based joint location and communication}
& BS, IRS, DL, multi-users & DL pilot
& Joint optimization of beamforming and subcarrier allocation for rate and location
& AO + SDR + MM \cite{CAS10_BoWang_TITS2024} \\
\hline

\parbox[t]{2.8cm}{\raggedright Beam sweeping to enhance sensing}
& BS, IRS, one user, one target & Echo of DL communication signal
& Adjust transmit beam to maximize sensing SNR based on echo and feedback
& Particle Swarm Optimization (PSO) \cite{SAC07_XYCao_TVT2023} \\
\hline

\multirow{3}{=}{\parbox[t]{2.8cm}{\raggedright Sensing via cooperative communication between users}}
& BS, mounted IRS, OFDM, multi-users & UL communication signals
& Decode UL data and estimate CSI parameters
& Bayesian inference, sum-product algorithm \cite{CAS08_XYuan_TWC2023} \\
\cline{2-5}
& Sphere IRS, no BS, communication between users & Pilot/data signal
& Estimate location parameters such as direction angles
& Subspace estimation \cite{CAS05_YXLin2025} \\
\cline{2-5}
& BS, IRS, communication between users & Pilot from BS and users
& Estimate location coordinates of mobile users
& Convex optimization \cite{CAS11_MMZhao_ISWCS2022} \\
\hline

\end{tabular}
\end{table*}

\subsection{C\&S Mutualism (CSM)}

Though not many, a number of existing works have investigated effective mechanisms to realize C\&S mutual benefit. Among the existing studies, one promising direction lies in exploiting communication data to strengthen sensing functions, whose outcome is, in return, to promote communication performance. The authors in \cite{SAC01_KTMeng_CL2023} proposed a C\&S mutual enhancement strategy in the V2I system, where IRS is mounted on each vehicle. The RSU provisions transmit downlink (DL) data to all vehicles in a time division multiple access (TDMA) manner. Each vehicle reflects other peer vehicles' communication signal back to the RSU in the time slot immediately before its own dedicated one, enabling the RSU to sense its angle direction through echo. This timely updated angle information is then utilized at the RSU to adjust beam direction to improve communication quality for subsequent DL data transmission. The work \cite{9966507,10197455} proposed SLAC protocol realizing C\&S mutual enhancement aided by three IRS devices, i.e., one primary passive IRS plus two semi-passive IRSs. Specifically, when a mobile user transmits uplink (UL) data, the primary IRS reflects the signal while the two semi-passive IRSs operate in sensing mode, extracting the AoA of the impinging signal to locate the user. The two semi-passive IRSs then turn into reflection mode, and the user's location will be utilized for beamforming adjustment to boost the subsequent communication efficiency. The reciprocal C\&S interplay scheme in \cite{9966507,10197455} was extended to multiple distributed IRS scenarios in \cite{locsensing1}. The authors of \cite{SAC09_XHu_TCom2024} divided the DL transmission of the IRS-aided wireless network into two stages. In the first stage, with IRSs' phase randomly set, the mobile user performs information recovery and initial AoA estimation through the received pilot and information symbols. Based on that, the IRSs' beams are aligned with the user's direction to promote the C\&S performance of the second stage transmission. In \cite{10475369}, the authors proposed a joint UL-DL reciprocal C\&S protocol empowered by two semi-passive IRSs. Concretely, during the UL stage, the IRSs work in sensing mode and extract users' AoAs through UL pilot and data symbols. The obtained AoAs are then utilized to configure IRSs' phase shifts in the DL transmission, where users further refine location estimation through DL pilot and information signals. The authors of \cite{CAS03_XTong_JSTSP2021} put forward a mutually benefiting C\&S scheme in a MU sparse code multiple access (SCMA) network to simultaneously implement uplink users' information recovering and environment sensing. By processing the uplink data in a sliding window manner and invoking the approximate message passing algorithm to exploit sparsity, the proposed scheme in \cite{CAS03_XTong_JSTSP2021} can progressively improve communication rate and environment sensing accuracy.

The authors of \cite{SAC17_JiguangHe_WCNC2020} conducted a joint codebook and scanning protocol design to realize simultaneous positioning and communication in an IRS-aided mmWave MIMO system. Specifically, the BS continuously transmits pilot signals and dynamically selects the IRS reflection pattern from a pre-designed codebook according to the user's successive feedback. As shown in \cite{SAC17_JiguangHe_WCNC2020}, the proposed hierarchical beam sweeping strategy yields a fast convergence to the mobile user's direction and remarkably improves the communication rate. The work \cite{Rc_YuHan_JSTSP2022} investigated the SLAC system deploying an extremely large IRS. By transmitting pilots and exploiting the NF spatially nonstationary characteristics of IRS cascaded channels, the authors of \cite{Rc_YuHan_JSTSP2022} developed methods that can accurately identify each user's visibility region (VR) and position. These sensing results, in return, provide more flexibility in scheduling communication users over non-overlapping VRs. The paper \cite{10521619} proposed a two-stage joint sensing and communication design in an IRS-aided system. During the sensing stage, the IRS reflection pattern forms the sensing code that can improve sensing accuracy. The sensed angle is then utilized to adjust the IRS phase shifts to boost spectral efficiency in the communication stage. The work \cite{HZhang_TVT2025} proposed an SLAC protocol for a network with distributed IRSs. The authors designed a space-time cooperative beam training process through the IRSs' dynamic reflection pattern to locate users. Based on the newly obtained location, a robust beamforming design is conducted to elevate the communication efficiency afterwards.

\begin{table*}[htbp]
\centering
\renewcommand{\arraystretch}{1.25}
\footnotesize
\caption{CSM Summary}
\begin{tabular}{|p{2.3cm}|p{3cm}|p{2.3cm}|p{4.8cm}|p{3cm}|}
\hline
\textbf{CSM Scheme} & \textbf{System Setup} & \textbf{Sensing Paradigm} & \textbf{Problems} & \textbf{Solutions} \\
\hline

\multirow{7}{=}{\parbox[t]{2.3cm}{\raggedright Communication data-aided sensing}}
& V2I, TDMA, DL, mounted IRS, multi-users
& Echo of DL communication signal
& Based on predictive AoA, maximize minimal communication rate
& Asymptotic analysis, AO, analytic solution \cite{SAC01_KTMeng_CL2023} \\
\cline{2-5}
& BS, UL, 1 passive IRS, 2 semi-passive IRSs, 1 user
& UL pilot + UL data
& Estimate AoAs, localization, beamforming to maximize communication SNR
& ESPRIT, AO, closed-form solution \cite{9966507} \\
\cline{2-5}
& BS, UL, 1 passive IRS, 2 semi-passive IRSs, 1 user
& UL pilot + UL data
& Estimate AoAs, localization, beamforming to maximize weighted sum of comm. and sensing SNR
& ESPRIT, AO, SDR \cite{10197455} \\
\cline{2-5}
& BS, UL, multi-passive/semi-passive IRSs, 1 user
& UL pilot + UL data
& Estimate AoAs, localization, beamforming to maximize communication SNR
& ESPRIT, AO, closed-form solution \cite{locsensing1} \\
\cline{2-5}
& BS, DL, 2 semi-passive IRSs, 1 user
& DL pilot + DL data
& Estimate AoAs, localization, beamforming to maximize communication SNR
& ESPRIT, analytic solution \cite{SAC09_XHu_TCom2024} \\
\cline{2-5}
& BS, DL, 2 semi-passive IRSs, multi-users
& UL data + pilot; DL data + pilot
& Estimate AoAs, localization; minimize auto/cross-beam correlation
& ESPRIT, AO, SDP \cite{10475369} \\
\cline{2-5}
& SCMA, IRS, UL, multi-users
& UL pilot
& Multi-user detection and environment sensing
& Compressive sensing (CS) via AMP algorithm \cite{CAS03_XTong_JSTSP2021} \\
\hline

\parbox[t]{2.3cm}{\raggedright Beam sweeping for location and communication}
& mmWave, BS, IRS, UL, 1 user
& UL pilot
& Design codebook and beam scanning protocol for IRS pattern
& Hierarchical dynamic scanning protocol for beam alignment \cite{SAC17_JiguangHe_WCNC2020} \\
\hline

\parbox[t]{2.3cm}{\raggedright Near-field localization and environment sensing}
& BS, large IRS, NF, multi-users
& UL pilot
& Identify users' VR, estimate location and multipath parameters
& Edge-based and accumulation-based VR detection; polar-domain hierarchical search \cite{Rc_YuHan_JSTSP2022} \\
\hline

\multirow{2}{=}{\parbox[t]{2.3cm}{\raggedright Joint design of sensing \& communication protocol and beamforming}}
& BS, IRS, one user
& DL pilot
& Optimize beamforming to minimize sensing CRB; Align beam to boost communication rate
& AO, convex optimization \cite{10521619} \\
\cline{2-5}
& multi-APs, multi-IRSs, multi-users
& DL pilot
& Design sensing scheduling protocol to minimize CRB; Optimize beamforming to maximize minimal communication rate
& GA for IRS selection, AO + SCA + gradient descent \cite{HZhang_TVT2025} \\
\hline

\end{tabular}
\end{table*}

\section{Networked ISAC Aided by IRS}
Achieving large-scale, high-accuracy sensing and localization in ISAC networks typically requires the collaboration of multiple anchor nodes or multiple BSs. Deploying IRSs as additional, hardware- and energy-efficient anchor nodes into the network is an appealing paradigm for enhancing networked sensing. IRS deployment facilitates high-accuracy positioning and improves joint sensing and communication performance in multi-BS cooperation scenarios. This section discusses the deployment of IRS for networked ISAC enhancement, including IRS-aided ISAC with one BS and IRS-aided ISAC network with multiple BSs.
\subsection{IRS-aided ISAC with One BS}
One appealing paradigm exploiting IRS to reinforce networked sensing is to deploy multiple IRSs as additional hardware and energy-efficient anchor nodes to facilitate high-accuracy positioning. A multi-IRS-aided positioning scheme generally has the following essential stages: i) measuring location-related parameters, e.g., AoA, AoD, time of arrival (ToA), and time difference of arrival (TDoA), associated with the LoS path with respect to BS (if applicable) and each separate IRS. Based on the measurements and geometry relationships between the anchors, location is initially obtained; ii) location is further refined by fusing all the measurements. For instance, the authors of \cite{Net14_WZhang_TWC2021} proposed to sequentially acquire every mobile device's AoA and AoD with respect to the BS and each individual IRS through beam training to fulfill coarse localization, which is further refined by the subsequent measurement fusion. As reported in \cite{Net14_WZhang_TWC2021}, the proposed multi-IRS-aided positioning achieves sub-meter level accuracy. The authors of \cite{Net03_GCAlexandr_WCL2022,Net06_JHe_ICASSP2023} and \cite{Net04_Acc2023_Zlatanov} proposed invoking the orthogonal matching pursuit (OMP) method, ANM combined with root MUSIC, and optimality analysis, respectively, to retrieve AoA/AoD in the first stage positioning, and conducting least-square (LS) fitting to all measurements in the second stage to achieve ML location estimation. In \cite{Net17_YifeiLiu_ICTC2023}, the authors investigated multi-IRS-aided positioning through first measuring both AoA and ToA with respect to each IRS through spatial smoothing and estimating signal parameters via rotational invariance techniques (ESPRIT) and then determining location via solving hyperbolic equations. The authors of \cite{actlocal} proposed a sequential positioning protocol interleaving each IRS measurement and dynamic beamforming adjustment to acquire a highly accurate location. Specifically, the authors employed long short-term memory (LSTM) units processing sequential measurements from each IRS and extracting temporal correlations into a fixed-dimension hidden state, which is fed into a deep neural network (DNN) to configure the IRS reflection pattern and BS beamforming for the subsequent pilot-based measurements. As shown in \cite{actlocal}, the proposed process learns to gradually focus the beampattern onto the mobile device and progressively refine positioning accuracy. The work \cite{jointBFFeng} analysed multi-IRS-assisted localization performance and derived localization CRB. Based on that, the authors of \cite{jointBFFeng} developed an AO procedure utilizing SDR and geometric programming techniques that effectively reduces transmit power while ensuring localization accuracy.

Besides positioning discussed above, multi-IRS augmented sensing networks can also implement diverse sensing applications, including velocity measurement, tracking, synchronization, simultaneous localization and mapping (SLAM), wireless fingerprinting recognition, and so on. The paper \cite{Net18_EZahra_SPL2022} proposed to measure the moving target's velocity based on a single-antenna pulse-Doppler radar station, which exploits echoes from multiple IRSs. By analysing the CRB of Doppler estimation, the authors of \cite{Net18_EZahra_SPL2022} designed IRS reflection patterns to improve velocity estimation accuracy. The paper \cite{Net15_XYuan_JSTSP2022} addressed the challenge of online location tracking in a downlink MIMO network deploying multiple IRSs. To effectively leverage temporal correlation, the authors established a probabilistic model of user mobility and conducted Bayesian inference on a factor graph to accomplish real-time tracking of mobile users. The authors of \cite{Net15_XYuan_JSTSP2022} further extended their study in a more sophisticated MIMO-OFDM context in \cite{Net19_XYuan_JSAC2023}, where the LoS path between IRSs may be temporarily obstructed. By modelling the LoS blockage as a birth-and-death process and utilizing variational Bayesian inference, \cite{Net19_XYuan_JSAC2023} developed a joint location tracking and blockage detection algorithm. Besides, in both \cite{Net15_XYuan_JSTSP2022} and \cite{Net19_XYuan_JSAC2023}, the authors evaluated tracking performance by deriving Bayesian Cram$\acute{\text{e}}$r-Rao bound (BCRB) and conducted IRS phase-shift optimization to improve tracking precision towards minimizing BCRB. The work \cite{Net22_WXu_JSPSP2022} investigated communication-aided timing synchronization and parameter estimation of cascaded IRS channels. A two-stage communication procedure is considered in \cite{Net22_WXu_JSPSP2022}. The BS transmits pilot signals to jointly estimate timing offsets and cascaded channels in the first stage, and then transmits information data along with applying timing offset equalization in the second stage. The authors of \cite{Net22_WXu_JSPSP2022} developed a majorization-minimization (MM) based algorithm, configuring the IRS phase shifts and timing offset equalizer to suppress timing synchronization error. The authors of \cite{Net05_HZhang_TWC2023} investigated an IRS-enhanced indoor SLAM paradigm, where a mobile agent (e.g., robot or drone) equipped with transceivers explores the environment. Deploying multiple IRSs on ceilings and walls, \cite{Net05_HZhang_TWC2023} proposed to optimize IRSs' phase shifts via genetic algorithm (GA) to elevate received signal strength, which remarkably improves multipath components (MPCs) sensing quality compared to the non-IRS case. The work \cite{Net07_Huang_CL2024} studied a multi-IRS empowered static environment sensing. Modelling ``environment'' as a sparse point cloud divided into small cubes, environment sensing is conducted by detecting the direction of the reflected signals from scatters. The authors of \cite{Net07_Huang_CL2024} proposed adjusting the IRSs' reflection rapidly within one coherence interval, which simulates a multichannel reception procedure, to determine the scatter distribution utilizing the ANM method. As demonstrated in \cite{Net07_Huang_CL2024}, deploying IRSs can achieve a much lower sensing error without any prior CSI. The work \cite{Net12_YLi_WCL2024} proposed a novel IRS-enabled indoor positioning scheme based on wireless fingerprinting using only one single-antenna BS. In \cite{Net12_YLi_WCL2024}, multiple IRSs are placed distributively around an indoor environment that is partitioned into mesh grids. To implement positioning, the distributed IRSs concurrently reflect the incident signal from the BS for consecutive time slots. The received signal sequence associated with each grid forms a specific pattern, namely the ``fingerprint'', to uniquely represent the grid. The authors of \cite{Net12_YLi_WCL2024} proposed maximizing the distinguishability among different grids' fingerprints by optimizing all IRSs' reflection pattern AO algorithm to realize accurate positioning. In \cite{Net09_PengChen_TSP2022}, the authors investigated IRS-enabled DoA estimation of impinging signals by a UAV swarm network. IRS elements are mounted on each UAV to reflect the impinging signal to the center UAV, which is equipped with one full receiving channel. The authors designed IRSs' phase shifts to obtain multiple measurements and extracted high-resolution AoD utilizing ANM and GD algorithms. 


The aforementioned papers focus on sensing capability. In comparison, a number of existing works consider dual C\&S functions in a single BS network augmented with multiple IRSs. e.g., \cite{Net21_JieXu_TCom2024,Net23_ZFei_IoTJ2024,Net20_DBao_TITS2025}. The paper \cite{Net21_JieXu_TCom2024} considered the ISAC scenario where BS intends to sense channel parameters (AoD or complete channel matrix) of multiple targets lying in the visual LoS area of IRSs, meanwhile communicating with mobile users. The authors developed algorithms jointly optimizing BS active beamforming and IRSs' phase shifts to minimize the worst CRB across all targets while ensuring the communication quality of communication users. The paper \cite{Net23_ZFei_IoTJ2024} investigated ISAC enhancement by utilizing multiple IRSs in the NLoS environment. The authors of \cite{Net23_ZFei_IoTJ2024} proposed deploying one hybrid IRS that contains active sensors besides passive reflecting elements to perform AoA sensing based on echoes and collaborative reflection beamforming, along with multiple passive IRSs. Adopting the BCD optimization framework and Riemannian gradient descent algorithms, \cite{Net23_ZFei_IoTJ2024} effectively optimizes the reflection pattern of IRSs to promote communication rate and sensing accuracy in a balanced manner. In \cite{Net20_DBao_TITS2025}, the authors investigated beamforming design for the DFRC system in the V2I context, where the IRS is mounted on each vehicle. The transmitted signal from the RSU is reflected by the IRSs cooperatively. The authors proposed to minimize the CRB of AoD estimation while ensuring the communication rate remains above a predefined level by optimizing IRSs' phase shifts via SDR optimization techniques.

Besides the above studies exploring various C\&S applications, one interesting aspect of networked ISAC lies in the placement of IRSs. The paper \cite{Net01_Nasri_WCL2022} investigated wireless localization with minimal infrastructure consisting of only one single-antenna BS and two IRSs. To conduct positioning through a three-step measurement process, the authors of \cite{Net01_Nasri_WCL2022} proposed to jointly optimize IRSs' phase shifts and placement, maximizing average receiving SNR to realize robust localization. The paper \cite{Net27_QQWu_TWC2024} analysed and compared the spectral efficiency between distributed and centralized deployment of multiple IRSs in a downlink communication network with one BS under various conditions. Specifically, in \cite{Net27_QQWu_TWC2024}, the distributed architecture divides the entire IRS elements into multiple small IRSs, each deployed near a cluster of users, as opposed to the centralized architecture with all IRS elements deployed near the BS. The analysis in \cite{Net27_QQWu_TWC2024} unveils that, for LoS or Rician fading channels, distributed IRS (spatial multiplexing gain) outperforms centralized IRS (passive beamforming gain) when the number of IRS elements is sufficiently large. 

\subsection{IRS-aided ISAC Network with Multiple BSs}
In this subsection, we discuss IRS-aided ISAC networks comprising multiple BSs. Firstly, a number of existing works have explored exploiting IRS in sensing networks with cooperative BSs to improve sensing capability. The paper \cite{Net16_XPang_TVT2023} investigated aerial target localization in half-duplex multi-BS networks deploying an IRS in the vicinity of each BS to eliminate the requirement for full-duplex operation of BSs. Selecting two BSs to respectively transmit and receive pilots, which is reflected by the target, and combining with the ON/OFF control of IRSs, the angular information of the target with respect to each BS is obtained. 
\begin{figure}
\centerline{\includegraphics[width=0.35\textwidth]{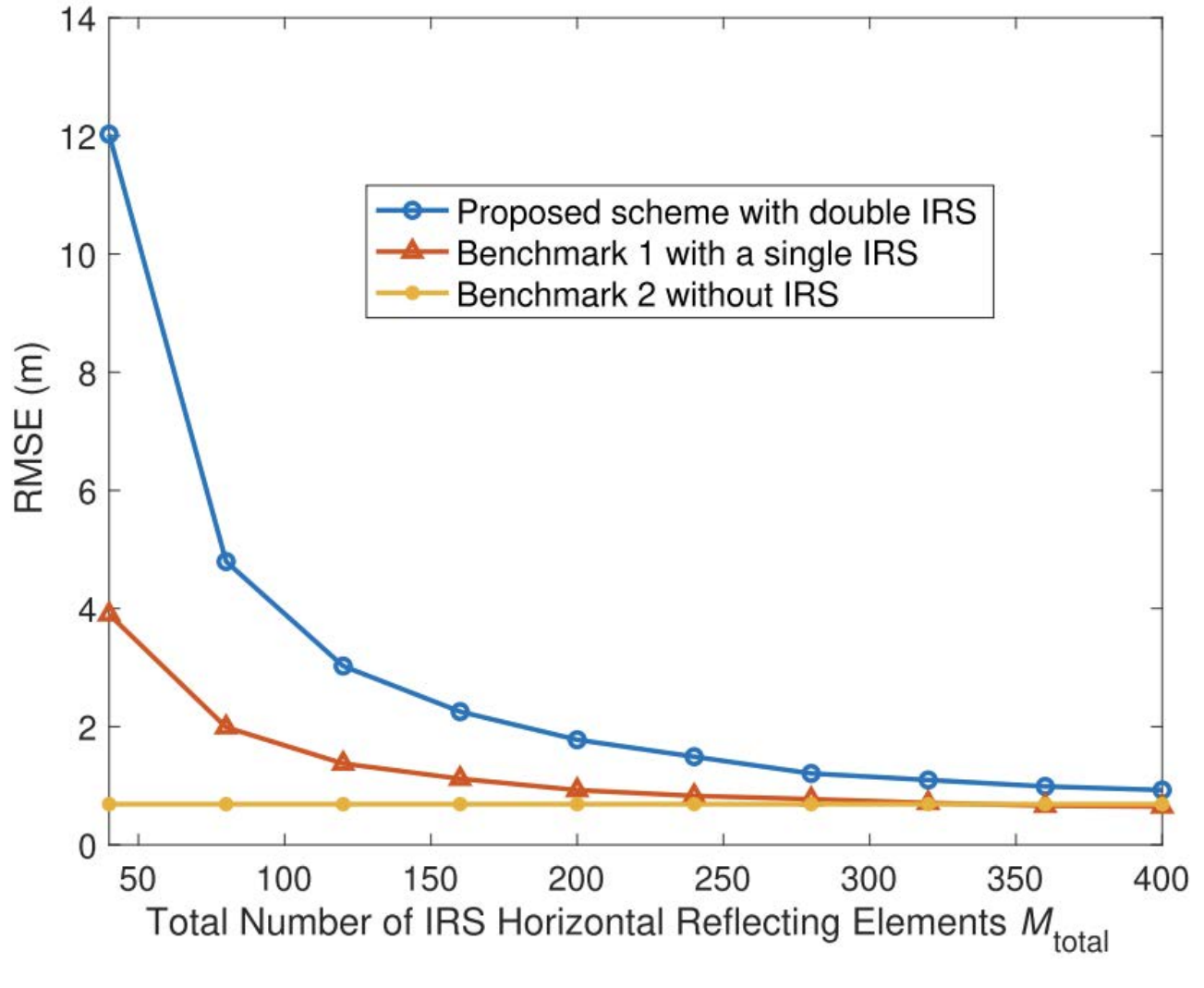}}
\vspace{-5pt}
\caption{RMSE versus total number of IRS horizontal reflecting elements.}	\label{network_isac}
\vspace{-10pt}
\end{figure}

To illustrate the role of cooperative IRSs in ISAC systems with multiple BSs, Fig. \ref{network_isac} shows simulation results for cellular sensing using cooperative IRSs . This scheme utilizes signals from an active BS reflected by a target, received by an idle BS, and employs cooperative passive beam searching via IRSs near both BSs to acquire target angle information for localization. The simulation results in Fig. \ref{network_isac} compare the Root Mean Squared Error (RMSE) of target localization versus the total number of IRS horizontal reflecting elements ($M_{total}$) for the proposed cooperative dual-IRS scheme (``Proposed scheme with double IRS''), a single-IRS benchmark (``Benchmark 1 with a single IRS''), and a no-IRS benchmark (``Benchmark 2 without IRS'') . It is observed that the proposed dual-IRS scheme significantly outperforms the two benchmarks, especially when $M_{total}$ is small . As $M_{total}$ increases, the RMSE for both the dual-IRS and single-IRS schemes decreases significantly, owing to the higher passive beamforming gain and finer angular resolution provided by more elements . When $M_{total}$ is sufficiently large (e.g., $M_{total} \ge 280$), this scheme using half-duplex BSs can achieve performance comparable to benchmark schemes requiring full-duplex BSs . These results demonstrate that deploying cooperative IRSs in cellular networks can effectively enhance network-level sensing (localization) accuracy.

Based on that, the target's location is determined via LS fitting. The work \cite{Net24_LWu_EuCAP2023} studied a multi-BS localization network in obstacle-dense urban environments, where BS-user LoS links are blocked. The authors of \cite{Net24_LWu_EuCAP2023} proposed establishing visual LoS channels by dynamically selecting distributive BS-IRS links. Specifically, \cite{Net24_LWu_EuCAP2023} developed a novel recursive localization scheme by utilizing a geometry-based visibility metric and propagation graph model (PGM), which combines the prior knowledge of BS-IRS LoS link condition and the real-time IRS-user channel quality, to select the assisting BS-IRS link for location refinement. Validation in \cite{Net24_LWu_EuCAP2023} demonstrates that the proposed recursive IRS selection scheme converges fast to the ground-truth user's position. The authors of \cite{mountlocal} proposed a novel mounted-IRS enabling device-free sensing scheme to implement six dimension (6D), i.e., location and  orientation, estimation. The mounted IRS on target reflects the probing signal from TX to RXs. The authors of \cite{mountlocal} developed a two-stage estimation method. Specifically, in the first stage, utilizing tensor decomposition to obtain AoAs, which is followed by location and orientation estimation by solving two least-squares problems, respectively. As verified by \cite{mountlocal}, the proposed mounted-IRS enabling 6D parameter estimation scheme can achieve CRB approaching accuracy. The paper \cite{Net28_SJin_TWC2023} proposed exploiting IRS in a multi-BS mmWave system to benefit user localization and environment sensing via enriching NLoS component measurements. Specifically, the authors proposed a two-stage sensing scheme. In the first stage, linear weighted LS (WLS) estimation with a closed-form solution is conducted to obtain UE location and velocity and scatters' positions based on measurements of both LoS and NLoS components. In the second stage, estimation of scatters is refined through LS fitting using NLoS measurements. As demonstrated in \cite{Net28_SJin_TWC2023}, the proposed IRS-enabled NLoS measurements can significantly improve sensing accuracy.

Besides, there is also a multitude of works considering both C\& S functions for IRS empowered ISAC networks composed of multiple BSs. The paper \cite{Net26_CHuang_TWC2025} evaluated the benefit of deploying IRSs in a multi-cell mmWave network on a system level. Adopting the realistic assumptions of Discrete Fourier Transform (DFT) codebook beamforming and easy user association policy, i.e., connecting either the nearest LoS BS or the strongest visual LoS IRS, the authors of \cite{Net26_CHuang_TWC2025} established a comprehensive probabilistic model to depict the network's behaviour, taking into account the spatial distribution of network elements (i.e., BSs, IRSs, users) and probabilistic blockage due to movability. Based on that, utilizing stochastic geometry principles, the authors of \cite{Net26_CHuang_TWC2025} derived coverage probability and average communication/sensing rate. As demonstrated in \cite{Net26_CHuang_TWC2025}, deployment of IRSs increases joint coverage rate of ISAC performance with growth from $62\%$ to $97\%$. The paper \cite{Net29_Duong_IoTJ2024} studied deploying IRS to enhance SLAC performance in a multi-carrier system composed of cooperative BSs, vehicular users, and targets. The authors of \cite{Net29_Duong_IoTJ2024} proposed a novel hybrid DRL approach to jointly optimize collaborative beamforming of BSs and IRS and sub-carrier allocation to minimize the positioning CRB while guaranteeing the network's communication capability. The paper \cite{Net31_TQuek_JSAC2025} investigated ISAC waveform design in cognitive radio (CR) and ISAC networks, where a primary BS (PBS) and secondary BS (SBS) coexist. In \cite{Net31_TQuek_JSAC2025}, the authors proposed to jointly optimize the SS time interval, SBS, and IRS beamforming to improve sensing SINR of secondary targets while ensuring the SS performance and both primary and secondary user's communication quality. The paper \cite{Net32_YuZhang_WCL2023} investigated jointly optimizing sensing and communication functions in a cloud radio access network (C-RAN) with sensing targets lying in NLoS regions. The authors of \cite{Net32_YuZhang_WCL2023} proposed to deploy an active IRS to amplify and reflect the joint communication and sensing signals from multiple remote radio heads (RRHs) while satisfying the SINR requirement for the communication users and the capacity limit of fronthaul links. The deployment of active IRS can effectively balance sensing and communication performance. The \cite{Net33_YuZhang_WCL2024} proposed to optimize beamforming of collaborative RRHs along with multiple IRSs in an OFDM Cloud Radio Access Network (C-RAN) system to synthesize sensing beampattern into a pre-designed shape while ensuring mobile users' QoS requirements and respecting the fronthaul link capacity constraint. The authors of \cite{Net33_YuZhang_WCL2024} developed an AO-based algorithm utilizing successive convex approximation (SCA) and SDR techniques, which facilitates high-rate communication while mitigating sensing beampattern distortion. The paper \cite{10411853} considered the joint design of active beamforming of cooperating BSs and IRS phase shifts to implement ISAC in a coordinated multiple points‌ (CoMP) scenario. The authors of \cite{10411853} proposed minimizing total transmit power while guaranteeing data rate requirement of communication users  and sensing mutual information (MI) constraints for multiple targets. As demonstrated in \cite{10411853}, IRS-assisted cooperative CoMP beamforming yields remarkable gain in power saving.

\begin{table*}[htbp]
\centering
\renewcommand{\arraystretch}{1.25}
\footnotesize
\caption{Networked ISAC with Mutiple BSs}
\begin{tabular}{|p{1.3cm}|p{1.7cm}|p{2.2cm}|p{4.5cm}|p{5.5cm}|}
\hline
\textbf{Scenario} & \textbf{System Setup} & \textbf{Sensing Paradigm} & \textbf{Problems} & \textbf{Solution} \\
\hline

\multirow{8}{=}{\parbox[t]{1.3cm}{\raggedright Sensing functional}}
& multi-BSs, multi-IRSs, 1 target
& Bistatic
& Target localization using half-duplex BSs
& Performs bistatic sensing aided by IRS between to BSs each time, obtains angular information, determines location via LS  fitting \cite{Net16_XPang_TVT2023} \\
\cline{2-5}
& Multi-BSs, DL, multi-IRSs
& DL pilot
& Dynamically selects BSs and IRSs for positioning
& Proposes an IRS selection rule based on visibility condition and PGM \cite{Net24_LWu_EuCAP2023} \\
\cline{2-5}
& 1 TX, multi-RXs, mounted IRS on target
& DL pilot
& 6D positioning
& Stage-1: tensor-decomposition to acquire AoA; stage-2: LS fitting by GD and manifold optimization \cite{mountlocal} \\
\cline{2-5}
& Multi-BSs, 1 IRS, UL, 1 user, scatters
& UL pilot
& Conducts user localization and environment sensing
& Stage-1: analytically estimates users' and scatters' parameters based on LoS and NLoS measurements, stage-2: refine estimation through NLoS measurements conducting weighted LS fitting \cite{Net28_SJin_TWC2023} \\
\hline

\multirow{12}{=}{\parbox[t]{1.3cm}{\raggedright Dual C\&S functionals}}
& Multi-BSs, 1 IRS, DL, multi-targets\&users
& DL echo
& Adopt practical beamforming and association strategy assumptions; Establish probabilistic model
& Derived coverage probability and communication \& sensing rates based on stochastic geometry \cite{Net26_CHuang_TWC2025} \\
\cline{2-5}
& Multi-BSs, 1 IRS, DL, multi-targets\&users 
& DL echo
& Jointly optimizes sub-carrier allocation and beamforming of BSs \& IRS to minimize positioning CRB s.t. communication QoS
& Hybrid DRL \cite{Net29_Duong_IoTJ2024} \\
\cline{2-5}
& CR, 1 PBS, 1 SBS, DL, 1 IRS, multi-PUs \&SUs \&targets
& DL echo
& Jointly optimizes SS time allocation, SBS beamforming and IRS phase shifts to improve worst sensing SINR s.t. SS performance and users' QoS requirements
& BCD, SCA, Dinkelbach algorithm \cite{Net31_TQuek_JSAC2025} \\
\cline{2-5}
& C-RAN, DL, 1 act-IRSs, multi-targets\& users
& DL echo
& Jointly optimizes RRHs' and active IRS' waveform to minimize beampattern matching error while ensuring mobile users' and fronthaul capacity constraints
& AO, SDR \cite{Net32_YuZhang_WCL2023} \\
\cline{2-5}
& OFDM, C-RAN, DL, multi-IRSs \&targets \&users
& DL echo
& Jointly optimizes waveform of RRHs and IRSs to minimize beampattern matching error while ensuring mobile users' and fronthaul capacity constraints
& AO, SDR \cite{Net33_YuZhang_WCL2024} \\
\cline{2-5}
& CoMP, 1 IRS, DL, multi-targets\&users
& DL echo
& Designs BSs beamforming and IRS reflection pattern to minimize transmit power s.t. sensing MI and communication rate constraints
& AO, SCA, CCP, \cite{10411853} \\
\cline{2-5}
\hline

\end{tabular}
\end{table*}

\section{Future Directions}
Although IRS-enabled sensing and ISAC systems show immense potential, there are several open research problems that are worthy of further investigation. This section presents several future research directions, broadly categorized into IRS-aided Sensing and IRS-aided ISAC. For the former, frontier topics such as six-dimensional movable IRS, mobile target sensing, and new sensing architectures will be explored. While for the latter, new opportunities and challenges in areas such as integrated sensing, communication, and powering (ISCAP), secure ISAC, satellite ISAC, near-field ISAC, and integration with movable antennas will be discussed.

\subsection{IRS-aided Sensing}
\subsubsection{Six-dimensional movable IRS} Most studies on IRS-enabled sensing assume fixed IRS platforms and ignore spatial DoF in surface placement and orientation. By integrating six-dimensional movable antenna (6DMA) technology \cite{6dmaCon,6dmaDis,6dmaMag,6dmaChan} with IRS, the six-dimensional movable IRS (6D-IRS) architecture can be obtained in which multiple passive IRSs adjust both 3D position and 3D rotation to maximize power gain and geometric gain according to target distribution. Joint optimization of IRS placement and orientation is crucial for enhancing localization accuracy and coverage; it also demands precise calibration of IRS position and orientation and tight synchronization with the BS. Although works on 6D‑IRS communication \cite{passive} and on 6DMA in sensing and communication \cite{10891142,6dmaSens,6dmaPar,shao2025tutorial,hy6d} have been extensively studied, a systematic sensing strategy for 6D‑IRS remains undeveloped. Moreover, existing IRS optimization methods depend on instantaneous CSI, whereas 6D-IRS requires CSI over a continuous space of candidate positions and rotations along with passive beamforming design, resulting in prohibitive computational complexity and pilot overhead. These challenges call for a rigorous mathematical formulation of IRS position and orientation design based on statistical CSI, together with efficient optimization algorithms.

\subsubsection{IRS for mobile target sensing} In practical scenarios, target mobility must be considered to maintain accurate localization. Omitting Doppler delay caused by movement degrades position estimates. Therefore, IRS-assisted target localization should account for target velocity and its effect on both location and orientation. Tracking mobile targets also requires NLoS channel identification and low-latency control of the IRS based on real-time target position feedback. Localization algorithms must adapt to propagation conditions that change rapidly. Moreover, IRS control hardware demands new designs. Current IRS phase update speeds are insufficient for high mobility cases. By the time the surface applies optimal phase shifts, the target has moved, and channel conditions have changed. Such delays can lead to outdated or ineffective configurations that harm sensing accuracy. Overcoming this challenge is essential for IRS application in fast-moving targets such as autonomous vehicles, UAVs, and high-speed trains \cite{10107972}.

\subsubsection{New IRS sensing architecture} Beyond the IRS sensing architectures discussed above, novel IRS structures offer significant potential and pose complex challenges. For example, a stacked IRS system comprises multiple metasurface layers that can be reconfigured to control electromagnetic wave propagation in multiple dimensions \cite{10534211}. Coupling between layers improves performance in multipath environments and enhances NF localization at mmWave and THz frequencies. The role of a stacked IRS in wireless sensing remains unexamined, and future work should develop analytical models for analog wave control to reduce reliance on digital signal processing. In addition, a clothing IRS employs thin and flexible antennas integrated into fabric to form a wearable sensing surface. By placing sensing elements across the garment, clothing IRS increases spatial sampling density and achieves finer detection and more accurate localization of moving targets. Integrating antennas into garments requires custom calibration for different shapes, which adds manufacturing complexity. Addressing these factors drives progress in wearable IRS sensing for dynamic scenarios.

\subsection{IRS-aided ISAC}
\subsubsection{IRS for integrated sensing, communication, and powering (ISCAP)} ISCAP will transform 6G networks into sustainable multi‑functional platforms \cite{chen2024isac}. Future research should explore how IRSs can enable efficient ISCAP operation by jointly optimizing signal sensing, information transmission, and energy delivery. IRS‑assisted beamforming can balance conflicting objectives such as estimation accuracy, data rate, and harvested power, but its dynamic coordination across devices and frequencies remains open. Moreover, networked IRSs could cooperatively manage interference, improve energy distribution, and enhance localization accuracy. Developing scalable optimization frameworks and AI‑driven control to achieve real‑time adaptation will be essential for practical ISCAP deployment in dense 6G environments.

\subsubsection{IRS for secure ISAC} Future secure ISAC systems must address vulnerabilities arising from the broadcast nature of joint sensing–communication signals. IRSs can dynamically reconfigure propagation environments to enhance physical‑layer security by strengthening legitimate links, steering artificial noise, and suppressing leakage toward eavesdroppers \cite{yu2020robust}. Leveraging sensing awareness also enables adaptive channel estimation for non‑cooperative targets and supports robust optimization under imperfect CSI. In addition, adjustable IRS reflection can confine or distort backscattered signals to preserve sensing privacy and mitigate jamming through path reconfiguration.

\subsubsection{IRS for satellite ISAC} Integrating IRSs into satellite ISAC systems offers great potential for extending global coverage and improving sensing–communication performance. IRSs can create virtual line‑of‑sight links to overcome blockages and adapt to rapidly varying satellite channels, enhancing reliability in dynamic environments \cite{ren2023robust}. Future work should address the unique challenges of space–terrestrial networks, including ultra‑precise synchronization, Doppler compensation, and interference management across orbital and terrestrial nodes. Developing adaptive beam tracking, interference‑aware control, and joint communication‑sensing waveform designs will be essential for realizing efficient and resilient IRS‑assisted satellite ISAC in 6G and beyond.

\subsubsection{IRS for near‑field ISAC} Beyond conventional far‑field designs, integrating IRSs into near‑field ISAC enables centimeter‑level resolution and high‑efficiency communication through precise beam focusing. Large‑aperture IRSs exploit spherical wavefronts to form tightly focused energy spots and achieve spatial multiplexing for concurrent sensing and communication. However, near‑field implementations face critical challenges, including accurate spherical‑wave channel modeling, severe beam‑training overhead, and hardware constraints such as phase quantization and coupling effects \cite{10663521}. Future research should develop low‑complexity beam training, learning‑based beam prediction, and joint waveform design to mitigate clutter and interference, paving the way for practical near‑field ISAC. 

\subsubsection{IRS with movable antennas for ISAC} Beyond conventional static metasurfaces, integrating movable antennas into IRS–aided ISAC systems unlocks dynamic environmental adaptability and precise beam control. By adjusting element positions and orientations in real time, movable IRSs can improve sensing resolution, maintain link reliability, and mitigate blockages in mobile or cluttered scenarios \cite{10777052}. However, practical realization requires efficient joint optimization of element movement and phase configuration, low‑latency control for large arrays, and robust modeling of time‑varying channels. Future research should develop lightweight control algorithms, accurate dynamic channel models, and vibration‑resilient hardware to enable agile IRS‑assisted ISAC.

\section{Conclusion}
This article has presented a review of IRS for enhancing wireless sensing and ISAC in 6G networks. It has been shown that IRSs can profoundly reshape wireless environments by introducing controllable reflections, thereby extending coverage, improving spatial resolution and sensing range, and enabling flexible trade‑offs between communication and sensing efficiency. From system design to network‑level optimization, IRSs offer significant additional DoFs for balancing coexistence, achieving mutualism, and enabling networked ISAC operations. Despite these advancements, several technical challenges remain open. Accurate modeling of high‑dimensional IRS‑assisted channels, low‑complexity joint active‑passive optimization, real‑time configuration in dynamic environments, and the integration of artificial intelligence (AI) for self‑adaptive control all require further exploration. To this end, IRS‑aided sensing and ISAC constitute a cornerstone technology for 6G by synergizing communication, sensing, and environmental intelligence to enable perceptive, energy‑efficient, and intelligent networks supporting immersive applications and pervasive connectivity.

\bibliographystyle{IEEEtran}  
\bibliography{references}      
\end{document}